\documentclass[prd,superscriptaddress,amsfonts,amssymb,amsmath,showpacs,twocolumn]{revtex4}
\usepackage{bm}
\usepackage{amsfonts}
\usepackage{latexsym}
\usepackage[latin1]{inputenc}
\usepackage{graphicx}
\usepackage{amsmath}
\usepackage{palatino}
\usepackage{mathpazo}
\usepackage{textcomp}
\usepackage{float}
\usepackage{booktabs}
\usepackage{dcolumn}
\usepackage{hyperref}
\hypersetup{colorlinks,citecolor=blue}
\usepackage{amsmath}
\usepackage{xcolor}
\usepackage{orcidlink}
\usepackage{mathtools}
\usepackage{subfig}

\def\jnl@style{\it}
\def\aaref@jnl#1{{\jnl@style#1}}

\def\aaref@jnl#1{{\jnl@style#1}}

\def\aj{\aaref@jnl{AJ}}                   
\def\apj{\aaref@jnl{ApJ}}                 
\def\apjl{\aaref@jnl{ApJ}}                
\def\apjs{\aaref@jnl{ApJS}}               
\def\apss{\aaref@jnl{Ap\&SS}}             
\def\aap{\aaref@jnl{A\&A}}                
\def\aapr{\aaref@jnl{A\&A~Rev.}}          
\def\aaps{\aaref@jnl{A\&AS}}              
\def\mnras{\aaref@jnl{Mon.~Not.~Roy.~Astron.~Soc.}}             
\def\prd{\aaref@jnl{Phys.~Rev.~D}}        
\def\prc{\aaref@jnl{Phys.~Rev.~C}}  
\def\prl{\aaref@jnl{Phys.~Rev.~Lett.}}    
\def\qjras{\aaref@jnl{QJRAS}}             
\def\skytel{\aaref@jnl{S\&T}}             
\def\ssr{\aaref@jnl{Space~Sci.~Rev.}}     
\def\zap{\aaref@jnl{ZAp}}                 
\def\nat{\aaref@jnl{Nature}}              
\def\aplett{\aaref@jnl{Astrophys.~Lett.}} 
\def\apspr{\aaref@jnl{Astrophys.~Space~Phys.~Res.}} 
\def\physrep{\aaref@jnl{Phys.~Rep.}}      
\def\physscr{\aaref@jnl{Phys.~Scr}}       
\def\commat{\aaref@jnl{Comm.~Math.~Phys.}}              
\def\science{\aaref@jnl{Science}}               
\def\cqg{\aaref@jnl{Classical Quant.~Grav.}}            
\def\jpcs{\aaref@jnl{JPCS}}                                     
\def\ijmpd{\aaref@jnl{Int.~J.~Mod.~Phys.~D}}                    
\def\grg{\aaref@jnl{Gen.~Relat.~Gravit.}}               
\def\rpp{\aaref@jnl{Rep.~Prog.~Phys.}}          
\def\npa{\aaref@jnl{Nucl.~Phys.~A}}        
\def\lrr{\aaref@jnl{Living Rev.~Rel.}}                   
\def\jcap{\aaref@jnl{J.~Cosmology Astropart.~Phys.}}    
\def\rmp{\aaref@jnl{Rev.~Mod.~Phys.}}   
\def\epjc{\aaref@jnl{Eur.~Phys.~J.~C}}

\allowdisplaybreaks[1]

\begin{document}
\color{red}

\title{Energy conditions in $f(Q,T)$ gravity}

\author{Simran Arora\orcidlink{0000-0003-0326-8945}}
 \email{dawrasimran27@gmail.com}
\affiliation{ Department of Mathematics, Birla Institute of Technology and Science-Pilani,\\ Hyderabad Campus,Hyderabad-500078, India}
\author{P.K. Sahoo\orcidlink{0000-0003-2130-8832}}
 \email{pksahoo@hyderabad.bits-pilani.ac.in}
\affiliation{ Department of Mathematics, Birla Institute of Technology and Science-Pilani,\\ Hyderabad Campus,Hyderabad-500078, India}
\date{\today}
\begin{abstract}
The recently proposed $f(Q, T)$ gravity (Xu et al. Eur. Phys. J. C \textbf{79} (2019) 708) is an extension of the symmetric teleparallel gravity. The gravitational action $L$ is given by an arbitrary function $f$ of the non-metricity $Q$ and the trace of the matter-energy momentum tensor $T$. In this paper, we examined the essence of some well prompted forms of $f(Q,T)$ gravity models i.e. $f(Q,T)= mQ+bT$ and $f(Q,T)= m Q^{n+1}+b T$ where $m$, $b$, and $n$ are model parameters. We have used the proposed deceleration parameter, which predicts both decelerated and accelerated phases of the Universe, with the transition redshift by recent observations and obtains energy density ($\rho$) and pressure ($p$) to study the various energy conditions for cosmological models. The equation of state parameter ($\omega\simeq -1$) in the present model also supports the accelerating behavior of the Universe. In both, the models, the null, weak, and dominant energy conditions are obeyed with violating strong energy conditions as per the present accelerated expansion.
\end{abstract}

\keywords{$f(Q,T)$ gravity; Energy Conditions; Deceleration parameter; Equation of state}
\pacs{04.50.Kd}
\maketitle

\section{Introduction}\label{sec1}

The recent and exciting advances of the cosmos are the accelerating expansion of the Universe \cite{Riess, Perlmutter, Spergel/2003, Spergel/2007}. A.G. Riess also mentioned that the dynamics of the ordinary matter present in the Universe are affected by more exotic forms of energy. Several phenomena have been accounted for, but according to the recent study, expanding illustration is considered as a consequence of a mysterious force dubbed dark energy (DE), which possesses a significant negative pressure. DE has the property that positive energy density and negative pressure satisfying $\rho+3p < 0$, which is known as the strong energy condition. Mainly there is two possible access to this expansion. First introduces some matter component on the right side of the Einstein equations such as the scalar field, cosmological constant, etc. These components acquire a negative equation of state (EoS) parameter. On the other, the second way is to modify the left side of Einstein field equations. It includes the modification in Einstein-Hilbert action by some arbitrary function $f $, which purely shows a geometric nature.

The $f(Q, T)$ gravity theory \cite{Yixin} is a recently proposed extension of symmetric teleparallel gravity in which the Lagrangian density is given by an arbitrary function of the non-metricity $Q$ and the trace of matter-energy momentum tensor $T$. The coupling between $ Q $ and $ T $ results in the non-conservation of the energy-momentum tensor, which entails some significant thermodynamic change of the Universe as in $f(R, T)$ gravity \cite{Harko}. The two different $f(Q, T)$ models in a flat FLRW space-time are studied with the proposed deceleration parameter. Further, the cosmic fluid adheres to the equation of state $p= \omega \rho$, where $\rho$, $p$, and $\omega$ represents the energy density, cosmological pressure, and EoS parameter.

The purpose of the present work is to study various energy conditions in the newly proposed $f(Q, T)$ gravity theory. In General Relativity (GR), energy conditions play a significant role in cosmology, black hole thermodynamics \cite{Hawking/1973}, singularity theorems \cite{Wald}. The energy conditions are the different ways through which one can effort to implement the idea of positiveness of the stress-energy tensor in the presence of matter and also observe the attractive nature of the gravity.  The energy conditions emerge from the Raychaudhuri equation in purely geometric nature with the provision that the gravity is attractive with energy density to be positive as well \cite{Santos}.

The Null, Weak, Strong, and Dominant energy conditions are the most fundamental energy conditions used in GR. The famous Hawking-Penrose singularity invokes the strong energy condition (SEC)  whose violation results in the observed accelerated expansion \cite{Hawking/1973, VISSER/2000}. The proof of the second law of black hole thermodynamics depends on the null energy condition (NEC) \cite{Wald, Visser}. On the other WEC is a combination of NEC ($\rho+p\geq 0$) and $\rho \geq 0$. This also means that the local energy density as measured by any timelike observer be positive. According to the definition, we can also see that if the NEC is violated, then to the descriptions of SEC, on the one hand, WEC and DEC on the other cannot be satisfied. We can consider the equation concentrating on a congruence of null geodesic and develop the null convergence condition $R_{\mu \nu}k^{\mu}k^{\nu}\geq 0$ for any null vector $k^{\mu}$. 

Energy conditions are studied in various modified gravity theories. S. Capozziello \cite{Capo} studied the energy conditions in GR using the power law in $f(R)$ gravity. K. Atazadeh \cite{K.Atazadeh} considered the effectiveness of the energy conditions in Brans-Dicke theory by conjuring the energy conditions from a generic $f(R)$ theory. Di Liu \cite{Liu} studied the energy conditions in $f(T)$ gravity by constraining this gravity with exponential $f(T)$ gravity, and Born-Infeld $f(T)$ gravity and M. Zubair \cite{Zubair} explored the validity of energy bounds in a modified theory of gravity which involves the non-minimal coupling of perfect fluid matter and torsion scalar. If we look at the work by T. Azizi \cite{Azizi}, he considered the energy conditions in the structure of the modified gravity, including the higher derivative torsional terms in action. Energy conditions in another modified gravity i.e., $f(G)$ gravity, have been discussed with different forms by N. M. Garcia \cite{Garcia} and K. Bamba \cite{Bamba}. S. Mandal\cite{S.Mandal} also discussed energy conditions in $f(Q)$ gravity. K. Atazadeh \cite{Atazadeh} also investigated  $f(R, G)$ gravity theory discussing the stability of cosmological solutions. F. G. Alvarenga \cite{Alvarenga} analyzed the perturbations and stability of de Sitter solutions and also the power-law solutions in $f(R, T)$ gravity. M. Sharif \cite{Sharif} introduced energy conditions in $f(G, T)$ gravity for two reconstructed models in the background of FLRW Universe. Viability of the bounds in higher-order derivative  $f(R,\square R, T)$ theory is investigated through energy conditions in \cite{Yousaf}.

The paper is organized in various sections. In section \ref{sec2} we briefly described $f(Q, T)$ gravity. In section \ref{sec3}, we introduced the proposed deceleration parameter. The energy conditions have been studied in section \ref{sec4}. In section \ref{sec5}, we have presented some $f(Q, T)$ models with various energy conditions. Finally, in section \ref{sec6}, we ended with the discussion and results.

\section{Overview of $f(Q, T)$ Gravity}\label{sec2}

The action in $f(Q,T)$ gravity is given as \cite{Yixin},

\begin{equation}  \label{1}
S=\int \left( \frac{1}{16\pi} f(Q,T) + L_{M}\right)d^4x \sqrt{-g} .
\end{equation}
where $f$ is an arbitrary function of the non-metricity $Q$ and the trace of the matter-energy-momentum tensor $T$, $L_{M}$ represents the matter Lagrangian and $g = det(g_{\mu\nu})$ and

\begin{equation}  \label{2}
Q\equiv -g^{\mu\nu}(L^{\alpha}_{\,\,\, \beta \mu} L^{\beta}_{\,\,\,\nu \alpha}-L^{\alpha}_{\,\,\,\beta\alpha}L^{\beta}_{\,\,\,\mu \nu}),
\end{equation}
where $L^{\alpha}_{\,\,\,\beta\gamma}$ is the deformation tensor given by, 
\begin{equation}  \label{3}
L^{\alpha}_{\,\,\,\beta\gamma}=-\frac{1}{2}g^{\alpha\lambda}(\nabla_{\gamma}g_{\beta\lambda}+\nabla_{\beta}g_{\lambda\gamma}-\nabla_{\lambda}g_{\beta\gamma}).
\end{equation}
\newline

The non-metricity $Q$ and trace of energy momentum tensor $T$ are defined respectively as 
\begin{equation}  \label{4}
Q_{\alpha} \equiv {Q_{\alpha}^{\,\,\, \mu}}_{\, \mu}, \hspace{0.15in} T_{\mu\nu}= -\frac{2}{\sqrt{-g}}\frac{\delta(\sqrt{-g}L_{M}}{\delta g^{\mu \nu}}
\end{equation}
and 
\begin{equation}  \label{5}
\Theta_{\mu \nu}= g^{\alpha \beta} \frac{\delta T_{\alpha \beta}}{\delta g^{\mu \nu}}.
\end{equation}

The variation of the gravitational action \eqref{1} leads to the following field equation

\begin{widetext}
\begin{equation}  \label{6}
8\pi T_{\mu \nu}= -\frac{2}{\sqrt{-g}}\nabla_{\alpha}(f_{Q}\sqrt{-g} P^{\alpha}_{\, \mu\nu}-\frac{1}{2}fg_{\mu \nu}+ f_{T}(T_{\mu \nu}+\Theta_{\mu \nu})-f_{Q}(P_{\mu\alpha\beta}Q_{\nu}^{\,\,\,\alpha \beta}-2Q^{\alpha \beta}_{\, \, \, \mu}P_{\alpha\beta\nu}).
\end{equation}
\end{widetext}
where $P^{\alpha}_{\,\,\, \mu \nu}$ is the superpotential of the model as mentioned in \cite{Yixin}.

We now assume a flat FLRW metric as, 
\begin{equation}  \label{7}
ds^{2}= -N^{2}(t)dt^{2}+ a(t)^{2}(dx^{2}+dy^{2}+dz^{2}),
\end{equation}
where $a(t)$ is the scale factor and $N(t)$ is the lapse function.The nonmetricity is $Q=  \frac{6 H^{2}}{N^{2}}$. The coincident gauge is set to make the relation trivial. So, in fixing the coincident gauge, we use the diffeomorphisms gauge equality. As we are not allowed to pick any specific lapse function, this is permitted by special cases in $Q$-theories, as $Q$ maintains a residual invariance to reparametrize time, which is well described in \cite{Beltran}. So, to use this type of  symmetry, $N(t)=1$ is chosen. Thus we have $Q=6H^{2}$ and the generalized Friedman equations are, 
\begin{equation}  \label{8}
8\pi \rho= \frac{f}{2}-6F H^{2}- \frac{2 \overline{G}}{1+\overline{G}}(\dot{F} H+F \dot{H}).
\end{equation}
and 
\begin{equation}  \label{9}
8 \pi p= -\frac{f}{2}+6F H^{2}+2(\dot{F}H+F\dot{H}) .
\end{equation}

where dot represents derivative with respect to time and $F= f_{Q}$ and $8 \pi \overline{G}=f_{T}$ represent differentiation with respect to $Q$ and $T$ respectively.

We can write the Einstein's field equations from Friedman equations as follows
\begin{equation}  \label{10}
3H^2=8\pi \rho_{eff}= \frac{f}{4F}-\frac{4\pi}{F} \left[(1+ \overline{G})\rho+\overline{G} p\right].
\end{equation}
and 
\begin{multline}  \label{11}
2\dot{H}+ 3H^2=-8 \pi p_{eff}= \frac{f}{4F}-\frac{2\dot{F}H}{F}\\
+\frac{4\pi}{F} \left[(1+ \overline{G})\rho+(2+\overline{G}) p\right].
\end{multline} 

\section{Cosmographic parameters}\label{sec3}

As we see that (\ref{10}) and (\ref{11}) are two equations with three unknowns viz $p, \rho$ and $H$. To find the exact solutions of these two equations, we need an additional physically viable condition. Here, we consider the deceleration parameter as studied in \cite{Campo}.  The deceleration parameter reads as 
\begin{equation} \label{12}
q(z)= -1+\frac{3}{2}\left( \dfrac{(1+z)^{q_{2}}}{q_{1}+(1+z)^{q_{2}}}\right) .
\end{equation}
where $q_{1}$ and $q_{2}$ must be positive definite. Further, these two parameters $q_{1}$ and $q_{2}$ are obtained from the observational constraints. The expression for Hubble's parameter $H(z)$ is constrained from the integrating the following equation,
\begin{equation} \label{13}
H(z)= H_{0}  exp\left[ \int_{0}^{z} (1+q(x))d(ln(1+x))\right] .
\end{equation}
 Introducing  (\ref{12}) into (\ref{13}), we obtained $H(z)$ as
 \begin{equation} \label{14}
 H(z)= H_{0} \left( \frac{q_{1}+(1+z)^{q_{2}}}{q_{1}+1}\right) ^{\frac{3}{2q_{2}}}.
 \end{equation}

It is well-known that the crucial quantity in explaining the evolution of the homogeneous and isotropic universe is the deceleration parameter defined as q. Its value lies in the fact that the rate at which the universe accelerates or decelerates its expansion. Although the universe is certainly accelerating, measurements of q still suffer from non-small uncertainties that are increasing with the higher redshifts. A cosmological model providing the expressions for $q(z)$ is of little help since none of them rests on the theoretical ground that is pinpoint convincing. So parametrized $q(z)$ based on practical and empirical reason can be useful unless observations aid some theoretical model. The proposed deceleration parameter containing two free parameters is valid from matter-dominated epoch $z\gg 1$ onwards i.e., up to $z=-1$. The functional form of q obeys $q(z\gg 1)= \frac{1}{2}$, which is very much demanded by cosmic structure formation. Also $q(z=-1)=-1$ required by thermodynamic arguments. Furthermore, the prediction by proposed $q(z)$ is consistent with spatially flat $\Lambda$CDM.
According to the parametrization studied in \cite{Campo}, the free parameters $q_{1}$ and $ q_{2} $ are fit with the observational data. The value is obtained as $q_{1}= 2.87^{+0.70}_{-0.53}$, $q_{2}=3.27^{+0.55}_{0.55}$, $H_{0}=70.5^{+1.5}_{-1.6}$. The $ q_{0} $ in this parametrization depends only on one free parameter i.e. $ q_{1} $. But if we take other parametrizations of $q$ discussed in \cite{Campo}, $ q_{0} $ depends on both free parameters $ q_{1} $ and $ q_{2} $. So, $ q_{0} $ in these cases results more degenerate than in the previous parametrization.

\begin{figure}[H]
\centering
\includegraphics[width=7.5 cm]{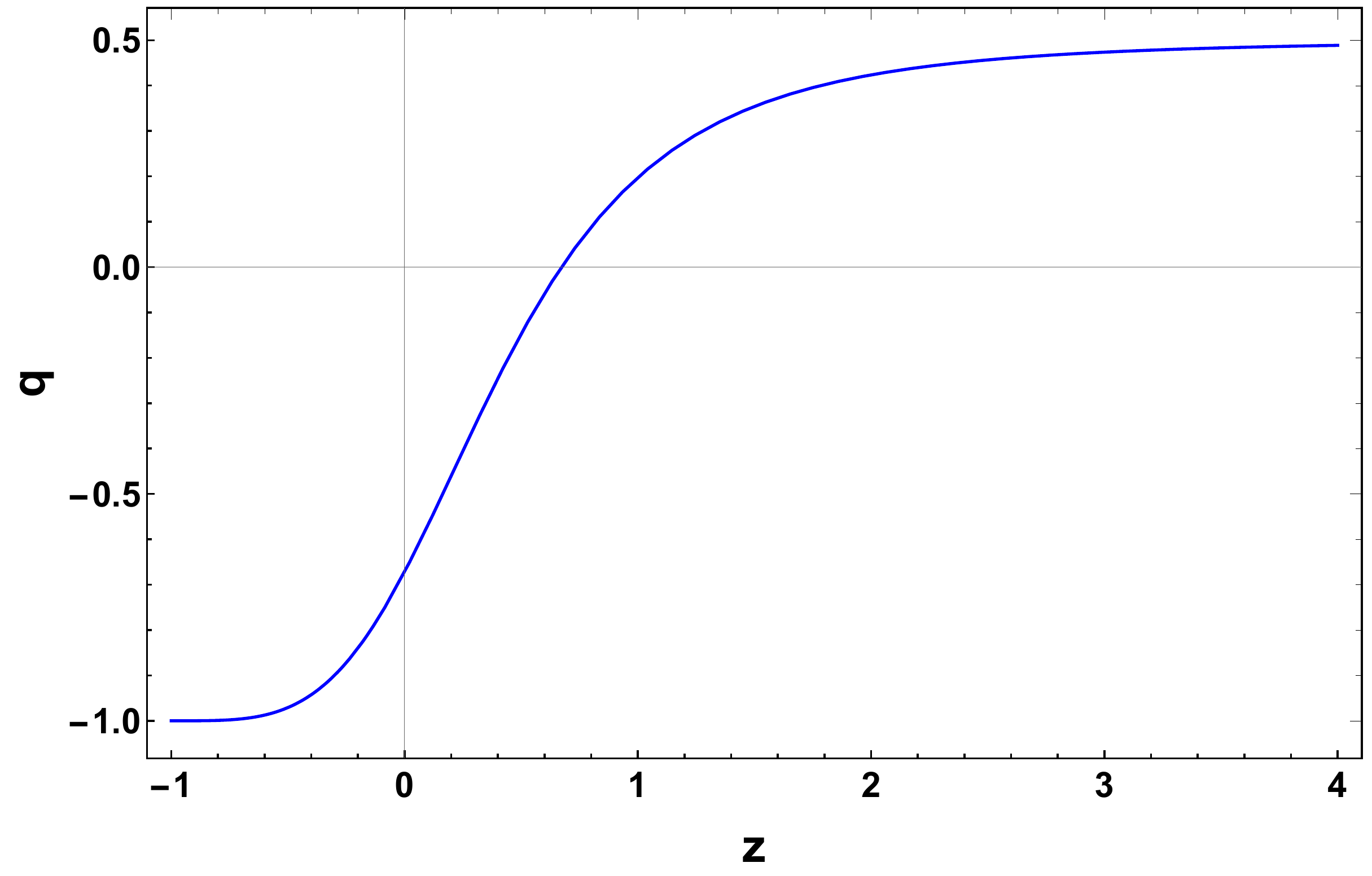}
\caption{Deceleration parameter versus redshift with values $q_{1}= 3.57$ and $q_{2}= 3.82$.}\label{fig1}
\end{figure}

The behavior of the deceleration parameter $q$ is presented in Fig.\ref{fig1}. One can detect the variation of $q$ from negative to positive at $z_{t}=0.67$ with $q_{0}=-0.67$. So the Universe exhibits the transition from early deceleration to the current acceleration. The results obtained for the transition i.e. redshift ($z_t$) and $q_0$ in the present model are consistent with values mentioned in literature \cite{Roman/2019,Davari/2018,Aviles/2017,Mamon/2017,Santos/2016,Gruber/2014,Aviles/2012,Cunha/2009,Ishida/2008}.

We try to study some more parameters in $f(Q, T)$ theory. If we expand the scale factor in Taylor series with respect to cosmic time \cite{Weinberg}, then this expansion points to a distance-redshift relation. In the Taylor series expansion, there comes the derivatives in the higher order of the deceleration parameter, which is known as a jerk($j$), snap($s$), lerk($l$) parameters. These parameters help us to perceive the past and future of the Universe. The sign of jerk parameter j regulates the change of the Universe's dynamics, a positive value indicating the instance of a transition time under which the Universe modifies the expansion. Furthermore, snap parameter s is essential to discriminate between an evolving dark energy term or behavior of cosmological constant.
According to the value of free parameters we have obtained $j_{0}=1.21$, $ s_{0}= -0.43 $ and $ l_{0}=3.6 $.
The exciting fact about the jerk parameter is, for $ \Lambda $CDM, the value of $j$ is one always. But according to our work, $j_{0}=1$ is dismissed. A similar argument has been given by S. Capozziello et al. in \cite{Capozziello}. Also, A. Aviles et al. \cite{Aviles} obtained results the same on these parameters.  

The three parameters are defined as follows \cite{Pan,Mandal},
\begin{equation} \label{15}
j= (1+z) \frac{dq}{dz}+ q(1+2q),
\end{equation}

\begin{equation} \label{16}
s= -(1+z) \frac{dj}{dz}- q(2+3q),
\end{equation}

\begin{equation} \label{17}
l= (1+z) \frac{ds}{dz}- s(3+4q).
\end{equation}

According to the considered deceleration parameter in \eqref{12}, the equations of jerk, snap and lerk reads as,

\begin{equation} \label{18}
j= \frac{2 q_{1}^2+q_{1} (3 q_{2}-5) (z+1)^{q_{2}}+2 (z+1)^{2 q_{2}}}{2 \left(q_{1}+(z+1)^{q_{2}}\right)^2}
\end{equation}

\begin{equation} \label{19}
s= \frac{\splitfrac{4 q_{1}^3-6 q_{1}^2 (q_{2}-2)^2 (z+1)^{q_{2}}+3 q_{1} \left(2 q_{2}^2-13 q_{2}+13\right)} {(z+1)^{2 q_{2}}-14 (z+1)^{3 q_{2}}}}{4 \left(q_{1}+(z+1)^{q_{2}}\right)^3}
\end{equation}

\begin{widetext}
\begin{equation} \label{20}
l= \frac{\splitfrac{4 q_{1}^4+2 q_{1}^3 \left(3 q_{2}^3-15 q_{2}^2+30 q_{2}-22\right) (z+1)^{q_{2}}-3 q_{1}^2 \left(8 q_{2}^3-54 q_{2}^2+95 q_{2}-53\right) (z+1)^{2 q_{2}}}{+ q_{1} \left(6 q_{2}^3-69 q_{2}^2+276 q_{2}-209\right) (z+1)^{3 q_{2}}+70 (z+1)^{4 q_{2}}}}{4 \left(q_{1}+(z+1)^{q_{2}}\right)^4}
\end{equation}
\end{widetext}
We have depicted the above parameters below.
\begin{figure}[H]
\centering
\includegraphics[width=7.5 cm]{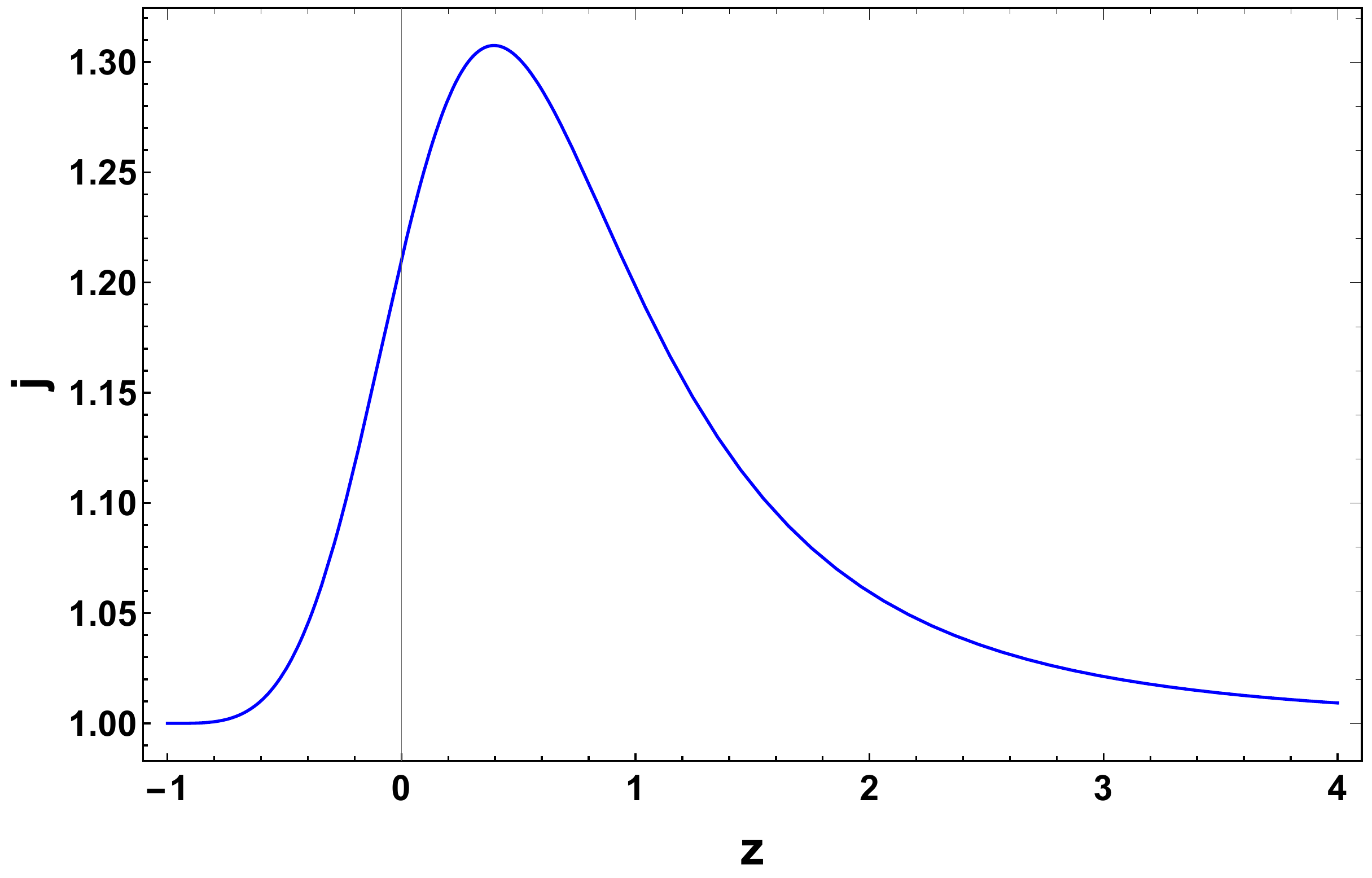}
\caption{Jerk parameter versus redshift with values $q_{1}= 3.57$ and $q_{2}= 3.82$.}\label{fig2}
\end{figure}

\begin{figure}[H]
\centering
\includegraphics[width=7.5 cm]{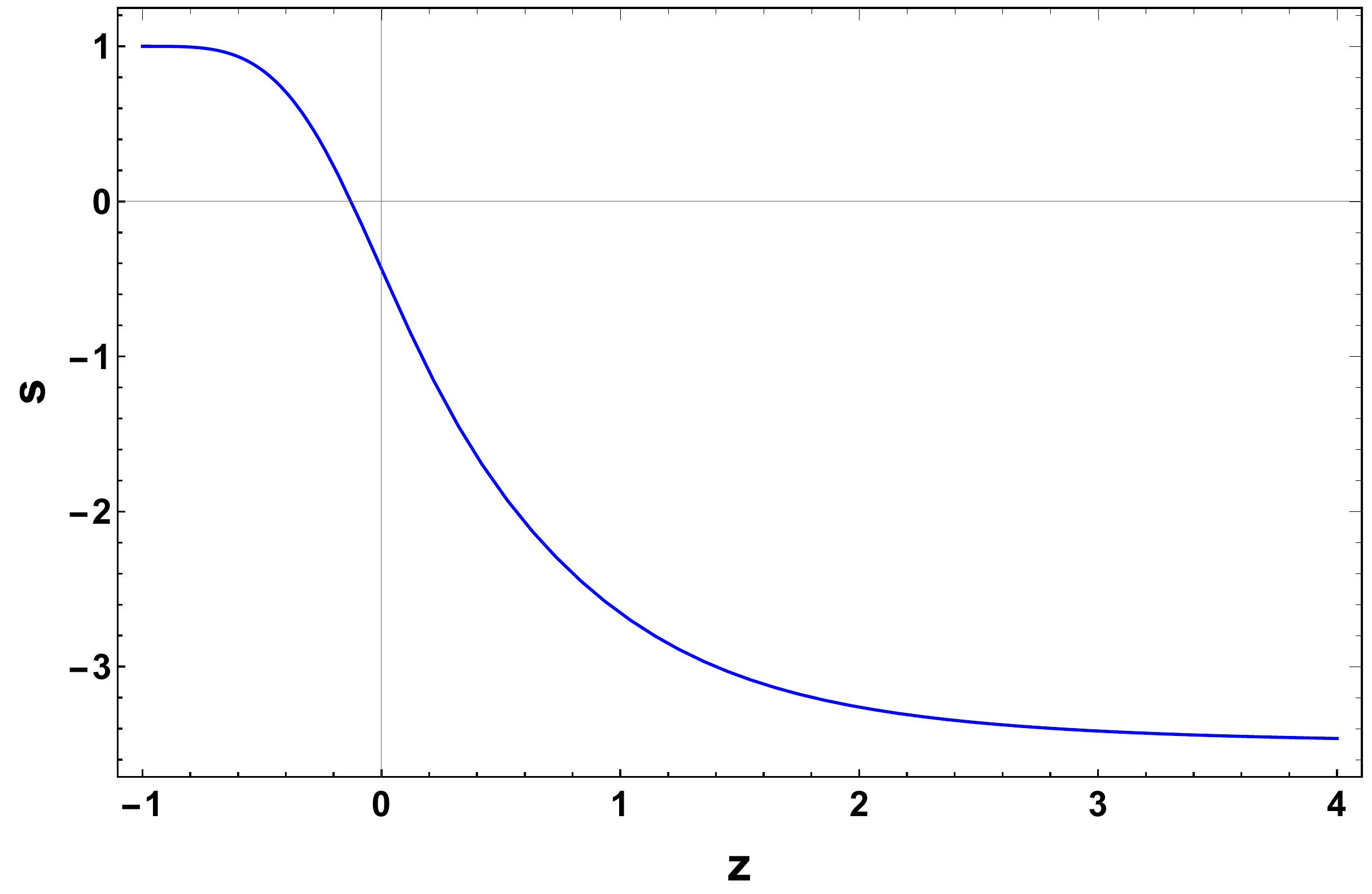}
\caption{Snap parameter versus redshift with values $q_{1}= 3.57$ and $q_{2}= 3.82$.}\label{fig3}
\end{figure}

\begin{figure}[H]
\centering
\includegraphics[width=7.5 cm]{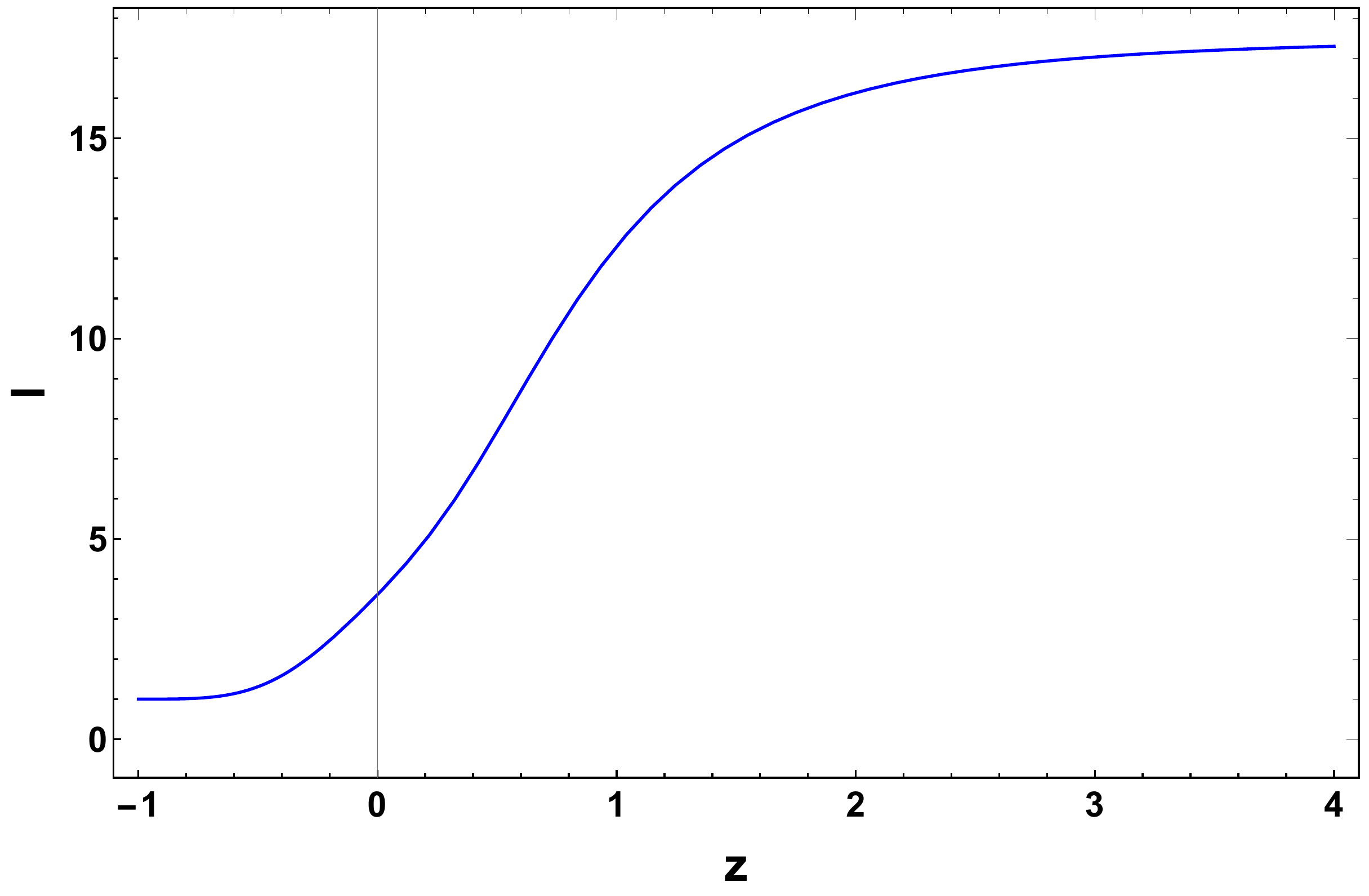}
\caption{Lerk parameter versus redshift with values $q_{1}= 3.57$ and $q_{2}= 3.82$.}\label{fig4}
\end{figure}

From Fig. \ref{fig2} and Fig. \ref{fig4} we can observe the positive behavior of jerk and lerk parameters representing the accelerated expansion. The jerk parameter shows the decreasing behavior whereas the lerk parameter increases. The snap parameter indicating negative behavior in Fig. \ref{fig3}  suggests the accelerated expansion. It shows that under certain modified gravity, the late-time acceleration can be seen in a strictly geometrical way.

\section{Energy Conditions} \label{sec4}

Shortly after the advent of general relativity, Weyl suggested an extension of Riemannian geometry in 1918, which he used for practical applications to develop the first unified theory of gravity and electromagnetism in which non-metricity of spacetime produced the electromagnetic field. Through incorporating an intrinsic vector field and semi-metric relation, Weyl generalized the Riemannian geometry \cite{Yixin/2020}. Also, torsion and non-metricity modify the typical Riemannian relationship between the metric tensor, the connection, and the space curvature. A non-Riemannian Geometry is a generalization of the normal Riemannian one when the space is equipped with torsion and non-metricity along with curvature as well. The structure of Raychaudhuri is by definition strictly geometrical. There has also been an attempt recently to extend the formulation of Raychadhuri to spaces with Weyl geometry \cite{Lobo/2015}.

Energy conditions are based on the Raychaudhuri equation, one of those conditions applied to the matter content in GR \cite{Kar/2007}. The Raychaudhuri equation gave emergence to these energy conditions with some requirements according to the attractive nature of gravity and also including the positive nature of energy density. These equations have a fundamental role in the coherence of null, timelike and lightlike geodesics. The Raychaudhuri equation \cite{Santos} in Riemann geometry reads,
\begin{equation}  \label{21}
\displaystyle \frac{d\theta}{d\tau}= -\frac{1}{3}\theta^{2}-\sigma_{\mu \nu} \sigma^{\mu \nu}+\omega_{\mu \nu} \omega^{\mu \nu}-R_{\mu \nu}u^{\mu}u^{\nu}.
\end{equation}
where $R_{\mu \nu}$ is the Ricci tensor and $\theta$, $\sigma_{\mu \nu}$ and $\omega_{\mu \nu}$ are expansion, shear and rotation associated to the vector field $u^{\mu}$ respectively.

From the above equation, the shear is a spatial tensor with $\sigma^{2}=\sigma_{\mu \nu} \sigma^{\mu \nu} \geqslant 0$. Thus the condition for attractive gravity for any hyper-surface orthogonal congruence lowered to 
\begin{equation}  \label{22}
R_{\mu \nu} u^{\mu} u^{\nu} \geqslant 0.
\end{equation}

Raychaudhuri equation with nonmetricity, in addition to curvature is given as \cite{Damianos/2018}
\begin{widetext}
\begin{multline}  \label{23}
\dot{\Theta} = -\frac{1}{3}\Theta^{2}-R_{\mu \nu} u^{\mu}u^{\nu}-2(\sigma^{2}-\omega^{2})+D^{\mu}a_{\mu}+\frac{1}{l^{2}}a_{\mu}A^{\mu}-\frac{1}{l^{2}}(a_{\mu}u^{\mu})-\frac{2\Theta}{3l^{2}}a_{\mu}u^{\mu}
+\frac{2}{3l^{4}}(a_{\mu}u^{\mu})^{2}+\frac{2}{l^{2}}a_{\mu}\xi^{\mu}- \dot{\overline{Q}}_{\mu}u^{\mu}\\
+\frac{1}{3}(\Theta+\frac{1}{l^{2}} a_{\nu}u^{\nu})(Q_{\mu}-\overline{Q}_{\mu})u^{\mu} - Q _{\mu \nu \lambda}(\sigma^{\mu \nu}+\omega^{\mu \nu})u^{\lambda}-\frac{1}{l^{2}}Q_{\mu \nu \lambda}u^{\mu}u^{\nu}(a^{\lambda}+\xi^{\lambda})+ Q_{\mu \nu \lambda}u^{\mu}\sigma^{\nu \lambda}\\
+\frac{1}{l^{2}}Q_{\mu \nu \lambda}(u^{\mu}\xi^{\nu}+a^{\mu}u^{\nu})u^{\lambda}+u^{\mu}u^{\nu}\nabla^{\lambda}Q_{\mu \nu \lambda}+Q_{\mu}^{\,\,\,\lambda \beta}Q_{\beta \lambda \nu}u^{\mu}u^{\nu}.
\end{multline}
\end{widetext}

where $\Theta= g^{\mu \nu} \nabla_{\mu} \nabla_{\nu}$ with $Q_{\mu \nu \lambda}=Q_{\mu (\nu \lambda)}$, $-l^{2}= u_{\mu}u^{\mu}$ , $A^{\mu}= u^{\lambda}\nabla_{\lambda}u^{\mu}$ and $a_{\mu}=u^{\lambda}\nabla_{\lambda}u_{\mu}$.

Now the energy conditions are defined as \cite{Mariam},
\begin{itemize}
\item Null energy condition (NEC) $\Leftrightarrow \rho + p \geq 0$.
\item Weak eneregy condition (WEC) $\Leftrightarrow
\rho + p \geq 0$ and $\rho \geq 0$.
\item Strong energy condition (SEC) $\Leftrightarrow \rho + p \geq 0$ and $\rho+ 3 p \geq 0$.
\item Dominant energy condition (DEC) $\Leftrightarrow \rho \geq |p|$ and $\rho \geq 0$.
\end{itemize}

\section{Cosmological $f(Q,T)$ Gravity Models}\label{sec5}

In this section we will investigate energy conditions for some specific models of $f(Q,T)$ gravity.

In the context of cosmology, the Hubble parameter $H$ and $q$ are related as,
\begin{equation} \label{24}
q= -\frac{1}{H^{2}} \frac{\ddot{a}}{a},
\end{equation}
Using (\ref{24}) we can refashion the equation as
\begin{equation} \label{25}
\dot{H}= -H^{2}(1+q).
\end{equation}

\subsection{$f(Q, T)= m Q + b T$}\label{A}

The simplest from of $f(Q,T)$ gravity is $f(Q,T)= m Q+ b T$ where $m$ and $b$ are model parameters. The values of $f_Q$ and $f_T$ in the field equation \eqref{6} are obtained as $F =f_{Q} = m$ and $8\pi \overline{G} = f_{T} = b$. Using the values of $\rho$ and $p$ from \eqref{10} \& \eqref{11} with $f=mQ+bT$ the energy conditions for this case are expressed as follows

\begin{widetext}
\begin{equation} \label{26}
NEC \Leftrightarrow \frac{m \left(b \left(6 H^2-2 H^2 (q+7)\right)-48 \pi  H^2\right)}{4 \left(b^2+12 \pi  b+32 \pi ^2\right)}-\frac{m \left(b \left(6 H^2 (q-1)+6 H^2\right)+8 \pi  \left(4 H^2 (q-2)+6 H^2\right)\right)}{4 \left(b^2+12 \pi  b+32 \pi ^2\right)} \geq 0.
\end{equation}

\begin{multline} \label{27}
WEC \Leftrightarrow \frac{m \left(b \left(6 H^2-2 H^2 (q+7)\right)-48 \pi  H^2\right)}{4 \left(b^2+12 \pi  b+32 \pi ^2\right)}-\frac{m \left(b \left(6 H^2 (q-1)+6 H^2\right)+8 \pi  \left(4 H^2 (q-2)+6 H^2\right)\right)}{4 \left(b^2+12 \pi  b+32 \pi ^2\right)}\geq 0 \\ 
\hspace{2.6mm} and \hspace{2.6mm}
\frac{m \left(b \left(6 H^2-2 H^2 (q+7)\right)-48 \pi  H^2\right)}{4 \left(b^2+12 \pi  b+32 \pi ^2\right)} \geq 0.
\end{multline}

\begin{multline} \label{28}
SEC \Leftrightarrow \frac{m \left(b \left(6 H^2-2 H^2 (q+7)\right)-48 \pi  H^2\right)}{4 \left(b^2+12 \pi  b+32 \pi ^2\right)}-\frac{m \left(b \left(6 H^2 (q-1)+6 H^2\right)+8 \pi  \left(4 H^2 (q-2)+6 H^2\right)\right)}{4 \left(b^2+12 \pi  b+32 \pi ^2\right)} \geq 0 \\ 
\hspace{2.9mm} and \hspace{2.9mm}
\frac{m \left(b \left(6 H^2-2 H^2 (q+7)\right)-48 \pi  H^2\right)}{4 \left(b^2+12 \pi  b+32 \pi ^2\right)}-\frac{3 m \left(b \left(6 H^2 (q-1)+6 H^2\right)+8 \pi  \left(4 H^2 (q-2)+6 H^2\right)\right)}{4 \left(b^2+12 \pi  b+32 \pi ^2\right)}\geq 0.
\end{multline}

\begin{multline} \label{29}
DEC \Leftrightarrow -\frac{m \left(b \left(6 H^2 (q-1)+6 H^2\right)+8 \pi  \left(4 H^2 (q-2)+6 H^2\right)\right)}{4 \left(b^2+12 \pi  b+32 \pi ^2\right)} \pm \frac{m \left(b \left(6 H^2-2 H^2 (q+7)\right)-48 \pi  H^2\right)}{4 \left(b^2+12 \pi  b+32 \pi ^2\right)} \geq 0 \\
\hspace{2.9mm} and \hspace{2.9mm}
\frac{m \left(b \left(6 H^2-2 H^2 (q+7)\right)-48 \pi  H^2\right)}{4 \left(b^2+12 \pi  b+32 \pi ^2\right)}  \geq 0.
\end{multline}
\end{widetext}

Here, we chose the model parameters as $m=-0.1$ and $b=59.1$ to make the energy density positive and the EoS parameter as per the observations.

\begin{figure}[H]
\centering
\includegraphics[width=7.5 cm]{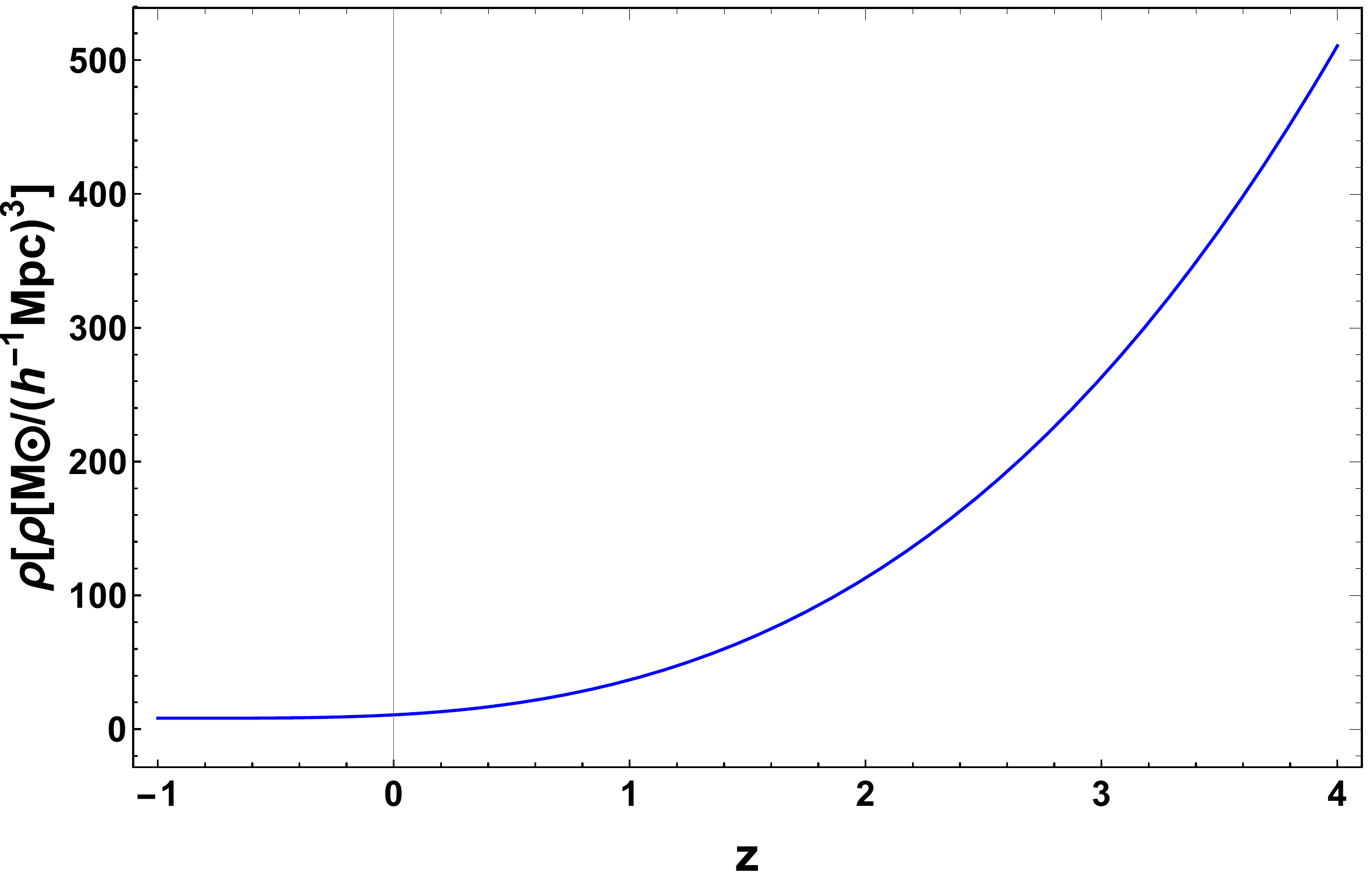}
\caption{Density parameter versus redshift.}\label{Fig-density}
\end{figure} 

Fig. \ref{Fig-density} portray the behavior of energy density versus redshift with appropriate choice of parameters as $m=-0.1, b=59.1, q_{1}=3.57, q_{2}=3.82$ and $H_{0}=68.9$. It can be observed that the energy density is always a positive function of redshift. At $z=0$, the energy density is strictly positive and increases with the increase in the value of $z$.

The EoS parameter is associated with energy density $\rho$ and pressure $p$. The EoS parameter classifies the expansion of the Universe. It represents the stiff fluid when $ \omega=1$. The matter dominated phase is represented by $\omega=0$ and if $\omega= \frac{1}{3}$, the radiation dominated phase is seen. Whereas the relation $-1< \omega \leq 0$ shows the quintessence phase and $\omega =-1$ shows the cosmological constant i.e., $\Lambda$CDM model. The phantom era is observed when $\omega<-1$.

\begin{figure}[H]
\centering
\includegraphics[width=7.5 cm]{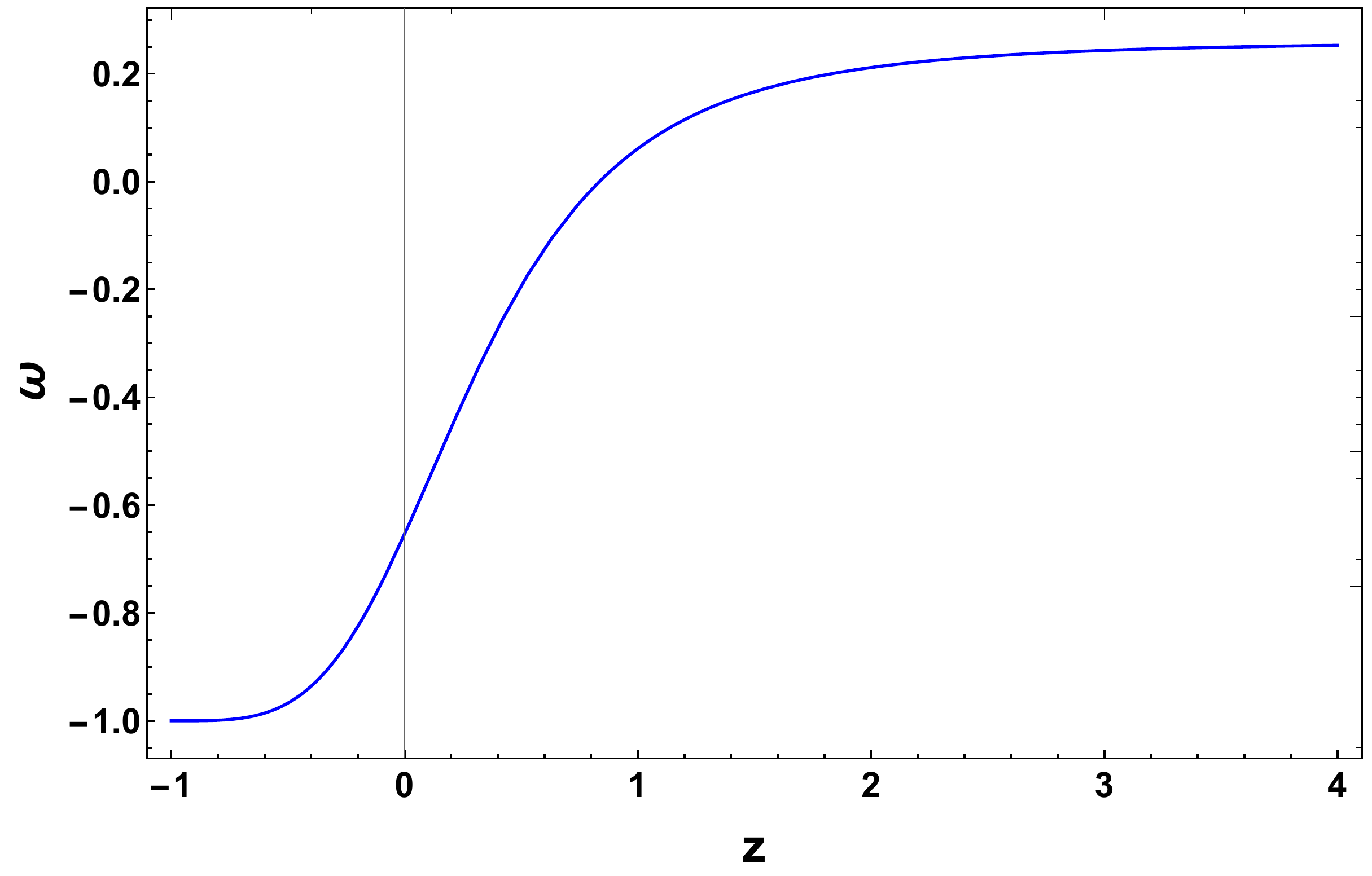}
\caption{EoS parameter versus redshift.}\label{Fig-Omega}
\end{figure} 

It is well known that for an accelerating Universe, we should have $\omega< -\frac{1}{3}$.  From Fig. \ref{Fig-Omega}, we can observe that the Universe exists the accelerating regime and enters the decelerated phase at an around $z= 0.835$ as studied in \cite{Capozziello}. The authors focused on the transition $z_{t}$ according to Planck 2015 data. He also stated the transition lies between 0.6 to 0.8. The little deviation occurs due to work done in modified gravity.
The energy conditions for this case are plotted below in Fig. \ref{Fig-EC1}.

\begin{figure}[H]
\centering
\includegraphics[width=7.5 cm]{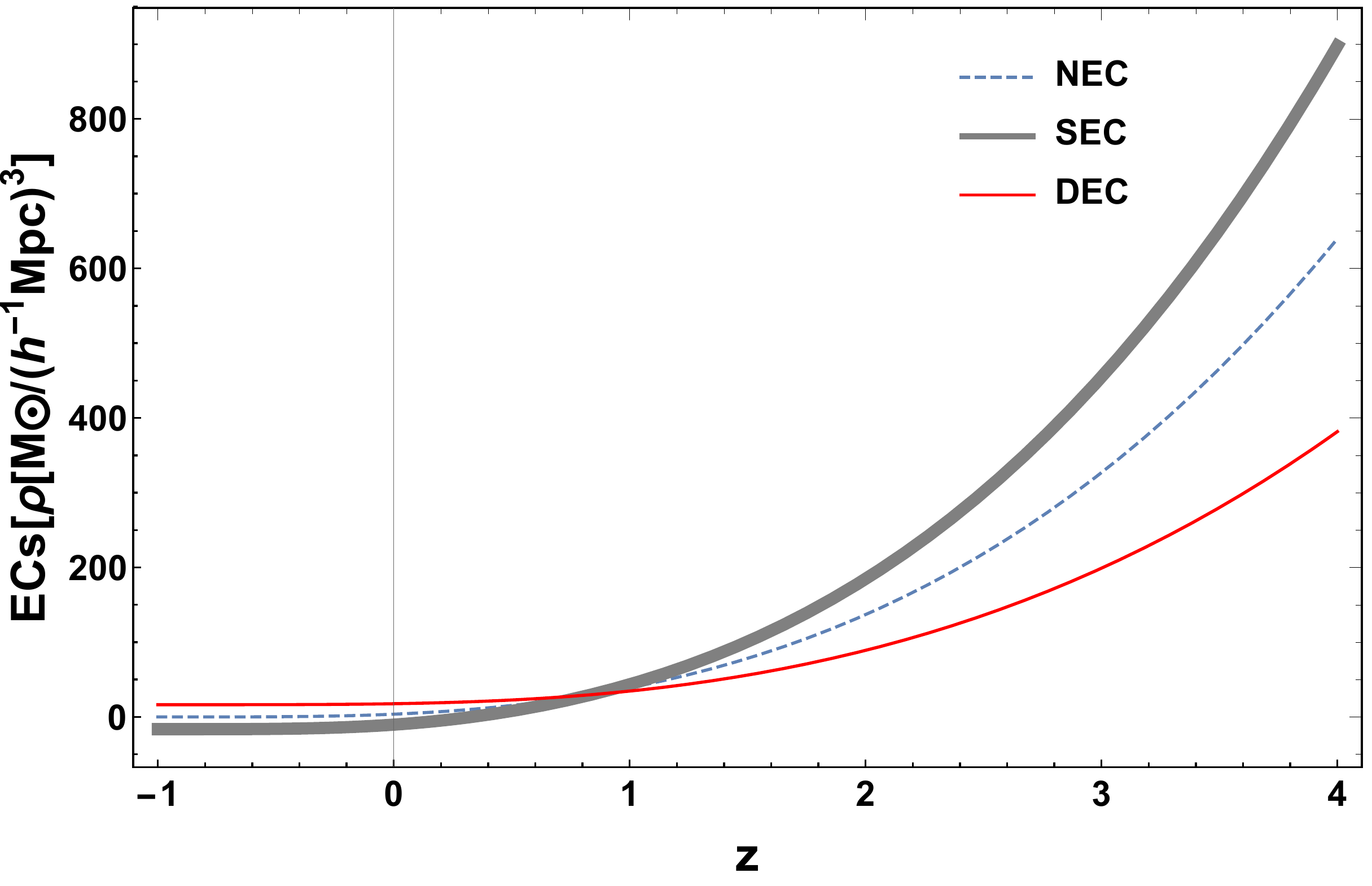}
\caption{Energy conditions versus redshift.} \label{Fig-EC1}
\end{figure}

\begin{widetext}
\begin{figure}[H]
  \centering
  \subfloat[variation of b]{\label{SECb}\includegraphics[width=80mm]{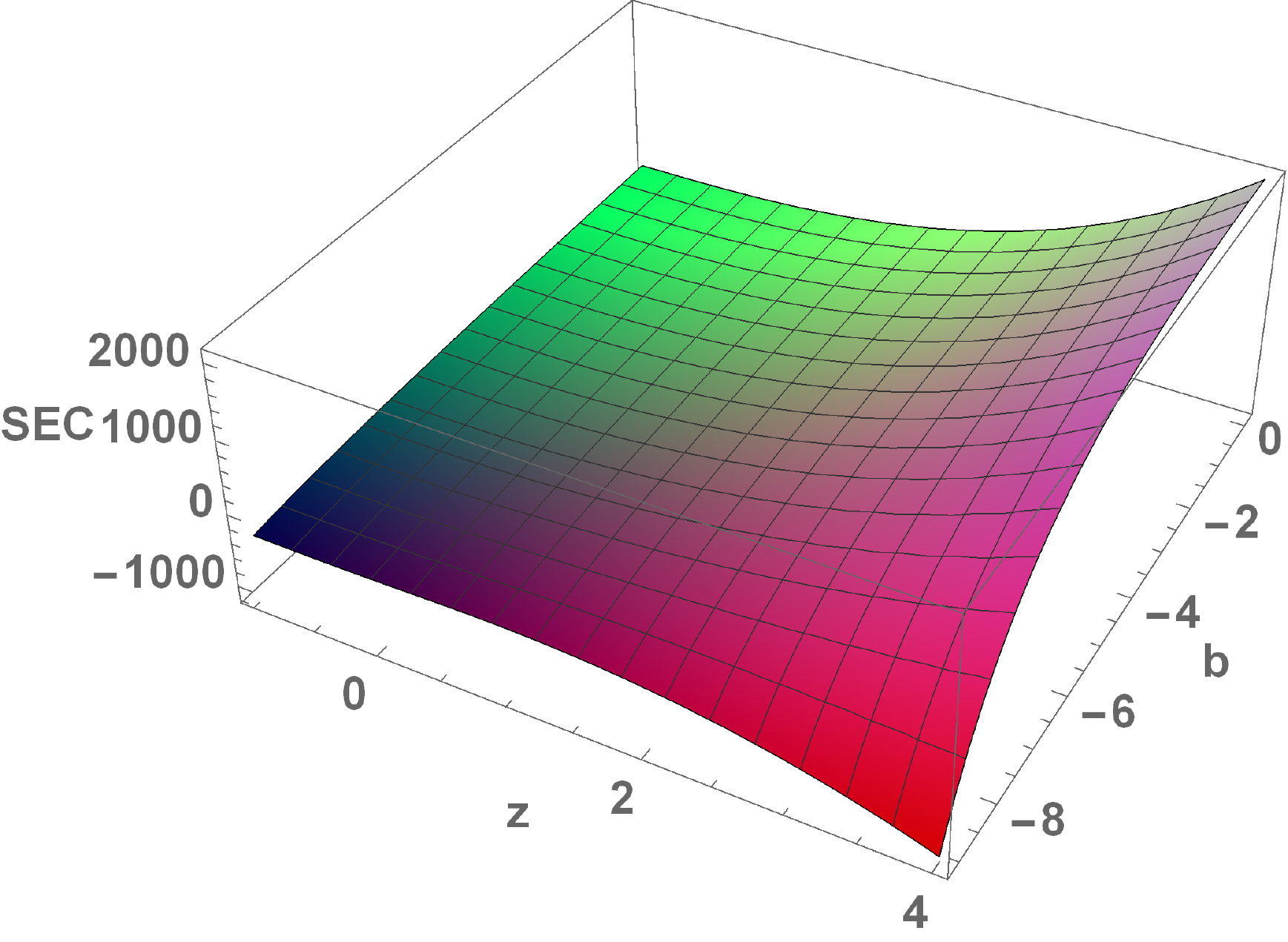}}
  \subfloat[variation of m]{\label{SECm}\includegraphics[width=80mm]{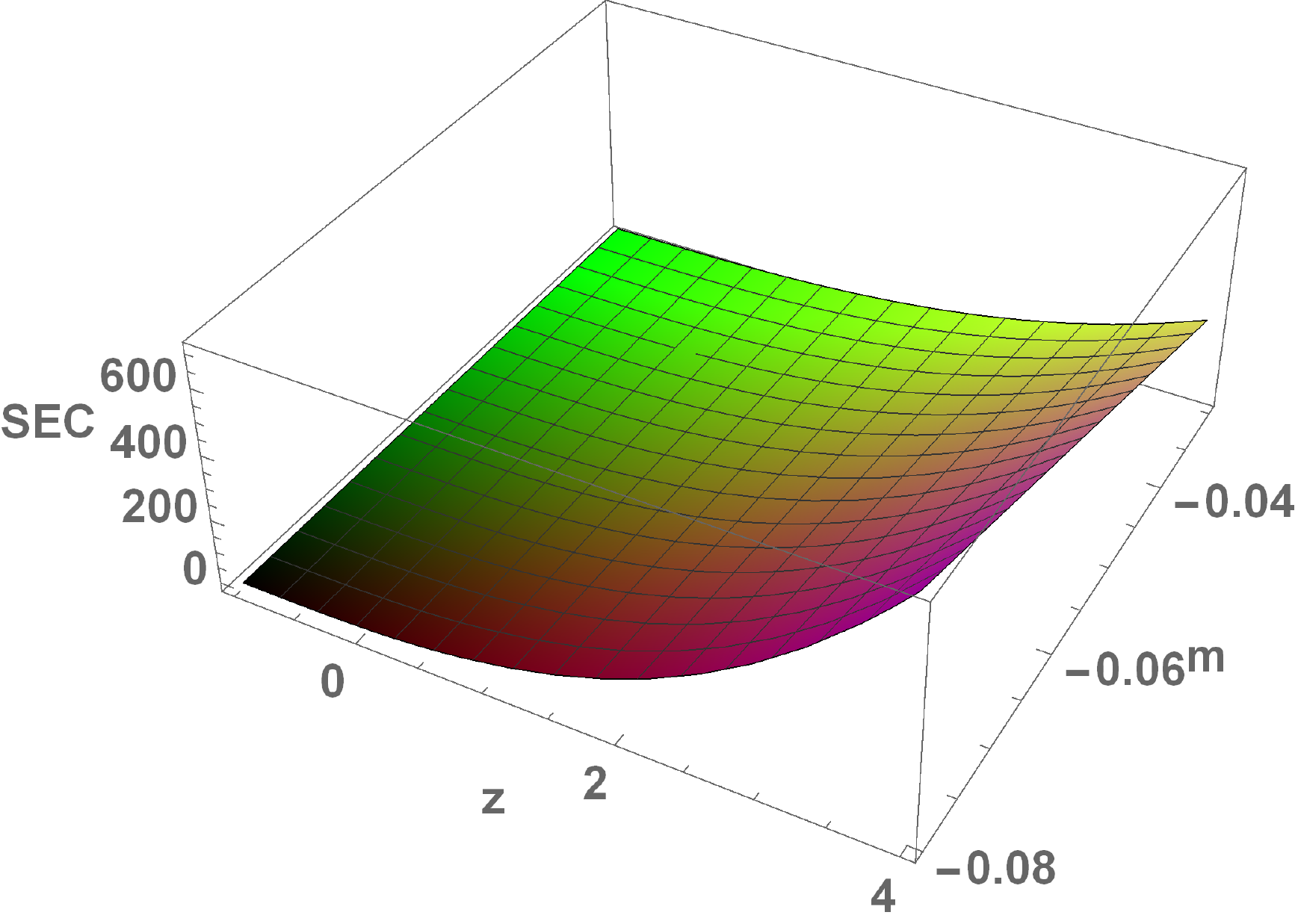}}
  \caption{ SEC with variation of b and m.}
\end{figure}
 
\begin{figure}[H]
  \centering 
  \subfloat[variation of b]{\label{NECb}\includegraphics[width=80mm]{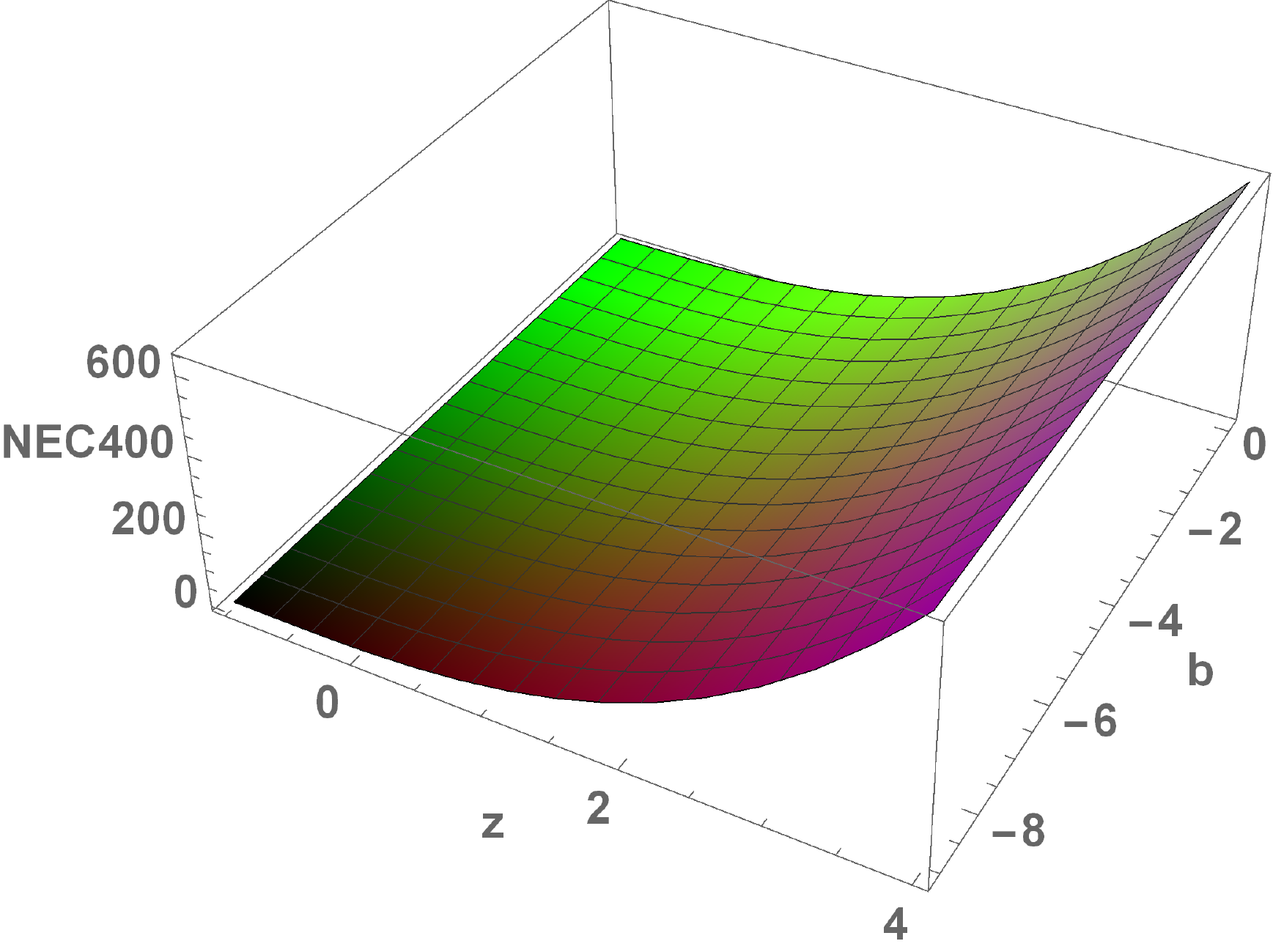}}
  \subfloat[variation of m]{\label{NECm}\includegraphics[width=80mm]{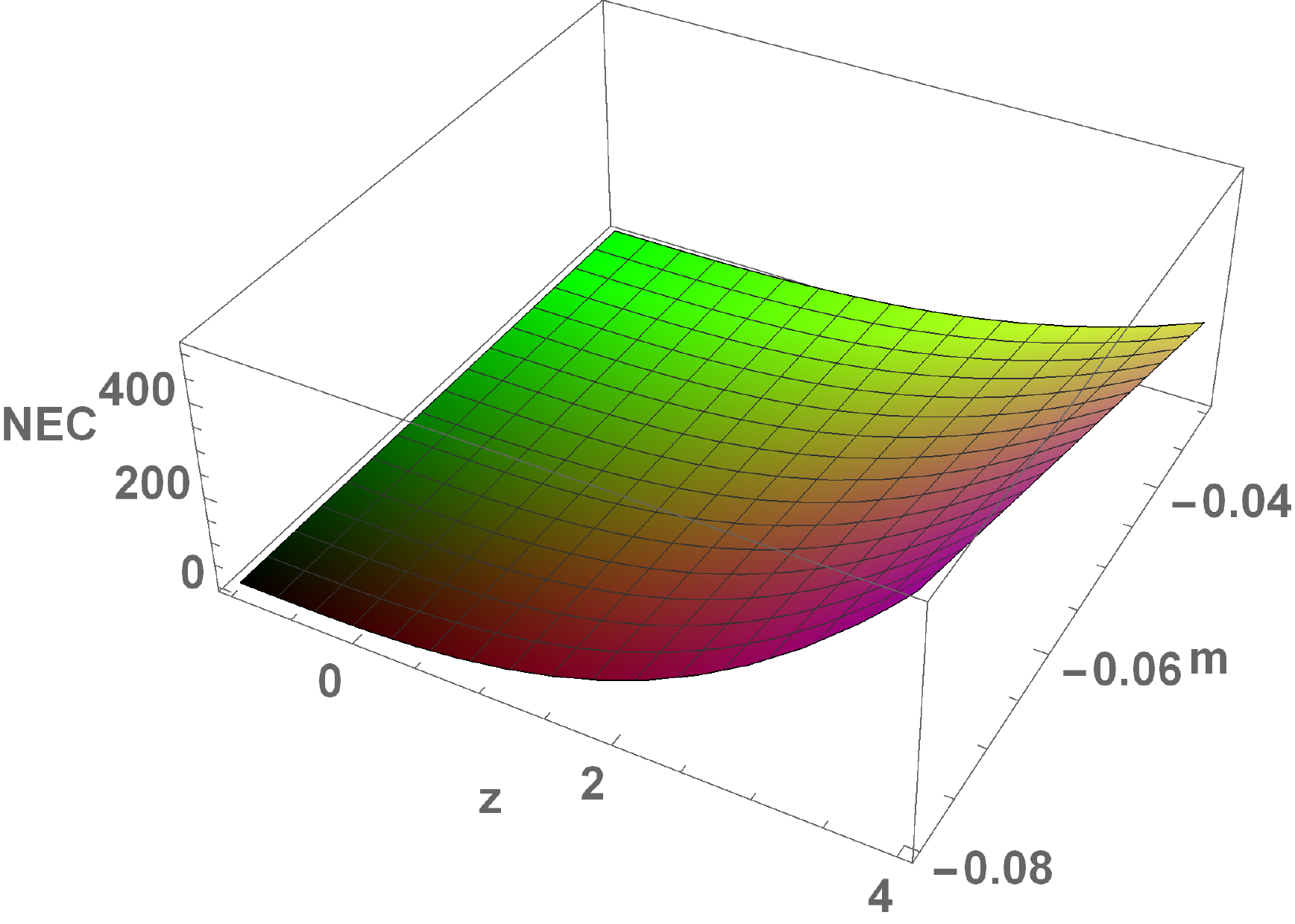}}
  \caption{ NEC with variation of b and m.}
 \end{figure} \label{NEC}

\begin{figure}[H]
  \centering 
  \subfloat[variation of b]{\label{DECb}\includegraphics[width=80mm]{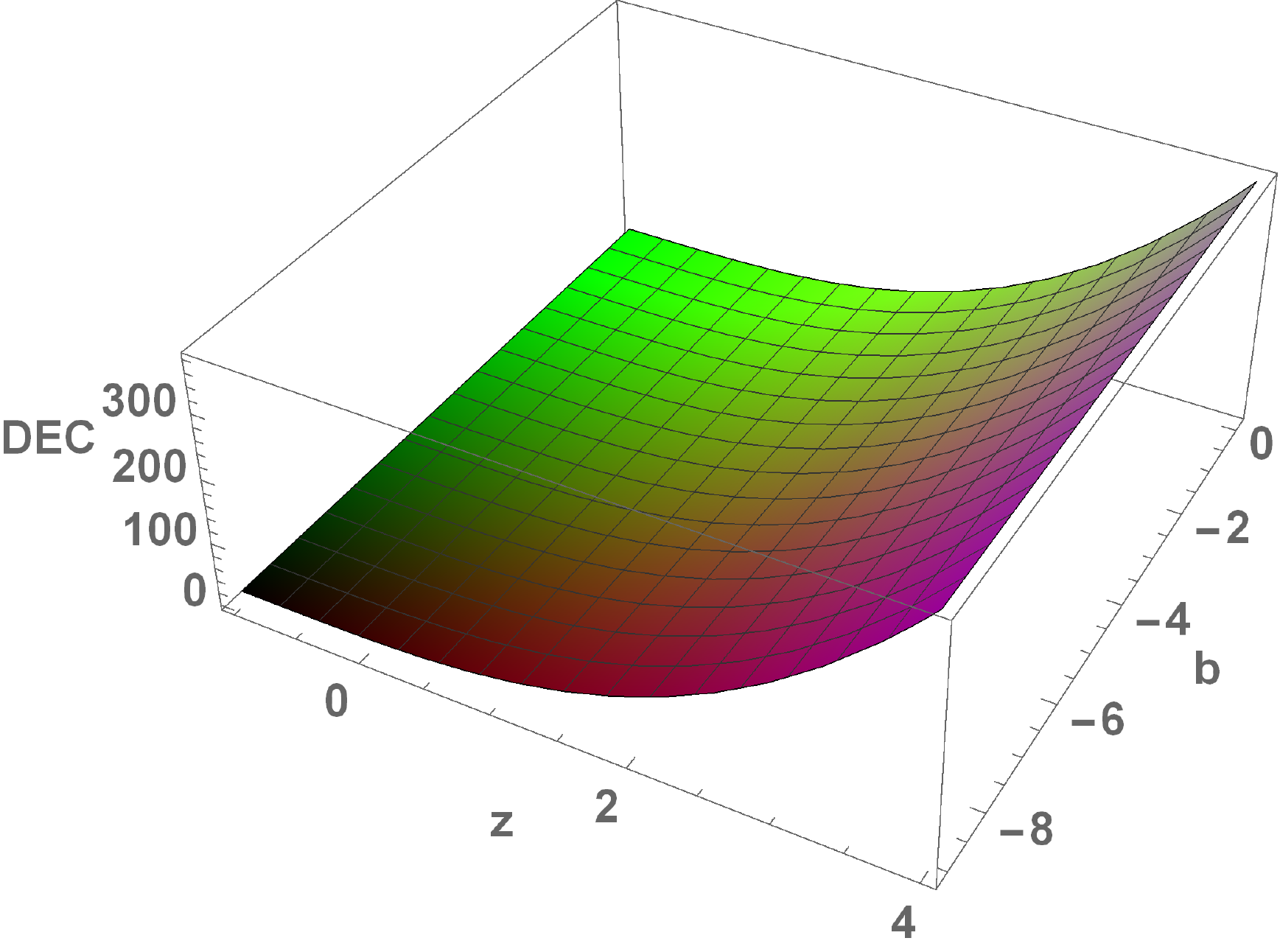}}
  \subfloat[variation of m]{\label{DECm}\includegraphics[width=80mm]{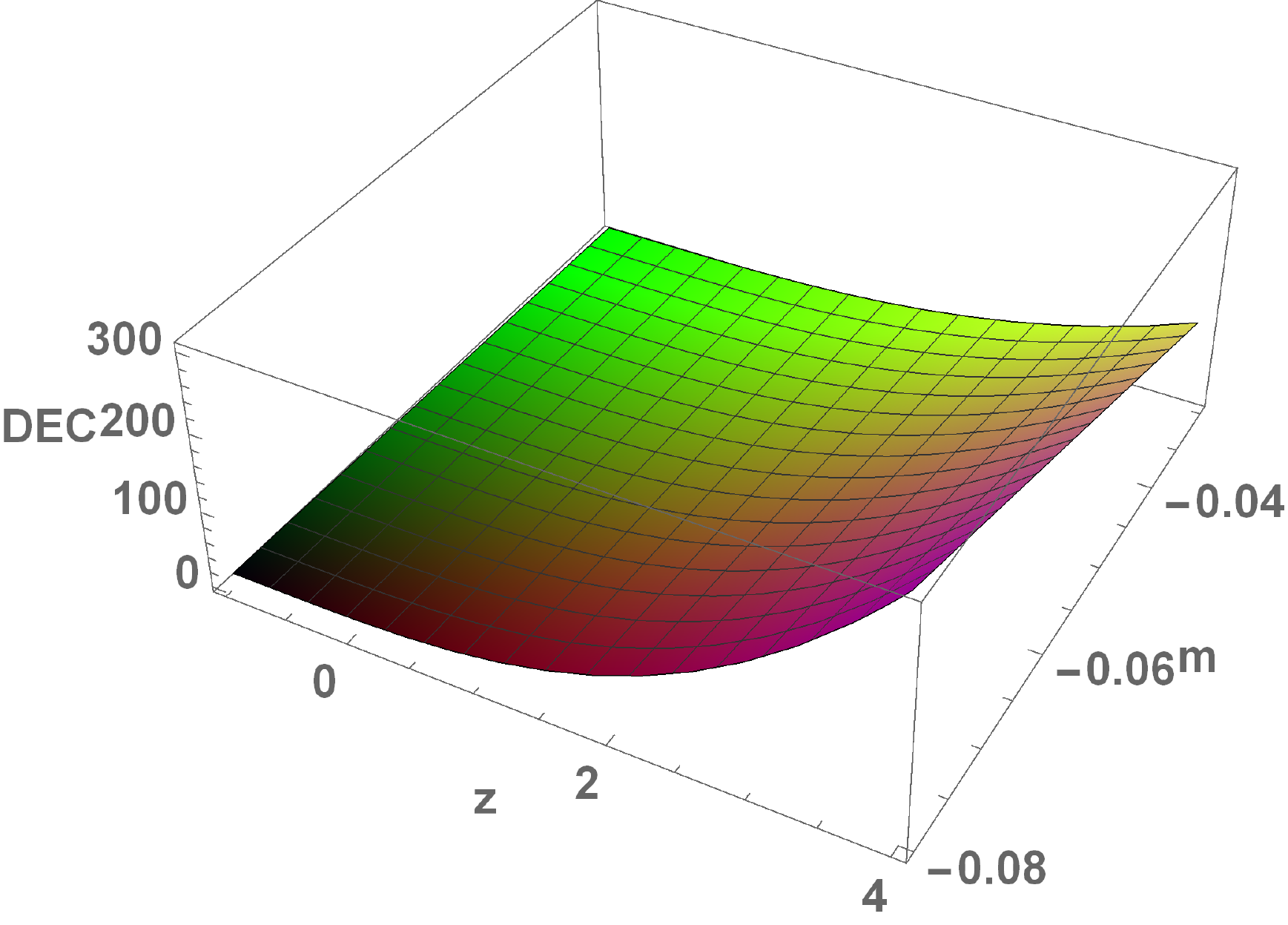}}
\caption{ DEC with variation of b and m.}
\end{figure} \label{DEC}
\end{widetext}

Among all the energy conditions, the strong energy condition is in the limelight of discussion. According to the recent data of the accelerating Universe, the SEC must be violated on cosmological scale \cite{Barcelo/2002, Moraes/2017}. Due to the SEC's importance, we observed the acceptable ranges of model parameters $b$ and $m$ in Fig. \ref{SECb} \& Fig. \ref{SECm}.  The variation of $m$ is from $-0.03$ to $-0.08$ whereas for $b$ it is $-9.1$ to $0.1$. The variation in $m$ and $b$ results in the variation of SEC behavior. In the case of variation of $b$, below -9.1, SEC shows some positive behavior. Also in case of $n$ from Fig. \ref{SECm} it is obeying.  We focused on the range when a violation is more. Also, $\omega$ negative implies $\rho+3p<0$. Therefore, there is a violation of the SEC at present. We also can see in Fig. \ref{Fig-EC1},\ref{NECb},\ref{NECm} and \ref{DECb}, \ref{DECm} that the NEC, and DEC are obeying.  Since we have shown the behavior of energy density in Fig. \ref{Fig-density}. We have also observed the behavior of NEC (i.e., partial condition of WEC). Therefore, validation of NEC and energy density together results in the validation of WEC.

\subsection{$f(Q, T)= m Q^{n+1}+b T$}\label{B}

In this case, we considered the second form of $f(Q, T)$ gravity as $f (Q, T )=m Q^{n+1} + b T$, where $m$, $n$ and $b$ are model parameters. Then we easily obtain
$F=(n+1)m Q^{n}$ and  $8\pi \overline{G}=b$. Here, the energy conditions are expressed as,

\begin{widetext}
\begin{equation} \label{30}
NEC \Leftrightarrow -\frac{m 2^{n+1} 3^n \left(2 n^2+3 n+1\right) (q+1) \left(H^2\right)^{n+1}}{b+8 \pi } \geq 0. 
\end{equation}

\begin{multline} \label{31}
WEC \Leftrightarrow -\frac{m 2^{n+1} 3^n \left(2 n^2+3 n+1\right) (q+1) \left(H^2\right)^{n+1}}{b+8 \pi } \geq 0 \\
\hspace{2.9mm} and \hspace{2.9mm}
-\frac{m 2^{n-1} 3^n (2 n+1) \left(H^2\right)^{n+1} (b (n q+n+q+4)+24 \pi )}{b^2+12 \pi  b+32 \pi ^2} \geq 0.
\end{multline}

\begin{multline}  \label{32}
SEC \Leftrightarrow  -\frac{m 2^{n+1} 3^n \left(2 n^2+3 n+1\right) (q+1) \left(H^2\right)^{n+1}}{b+8 \pi } \geq 0\\
\hspace{2.9mm} and \hspace{2.9mm}
-\frac{m 6^n (2 n+1) \left(H^2\right)^{n+1} (b (5 n (q+1)+5 q+2)+24 \pi  (n q+n+q))}{b^2+12 \pi  b+32 \pi ^2} \geq 0.
\end{multline}

\begin{multline} \label{33}
DEC \Leftrightarrow-\frac{m 2^{n-1} 3^n (2 n+1) \left(H^2\right)^{n+1} (b (n q+n+q+4)+24 \pi )}{b^2+12 \pi  b+32 \pi ^2} \mp \\
\frac{m 2^{n-1} 3^n (2 n+1) \left(H^2\right)^{n+1} (3 b (n q+n+q)+8 \pi  (2 n (q+1)+2 q-1))}{b^2+12 \pi  b+32 \pi ^2} \geq 0\\
\hspace{2.9mm} and \hspace{2.9mm}
-\frac{m 2^{n-1} 3^n (2 n+1) \left(H^2\right)^{n+1} (b (n q+n+q+4)+24 \pi )}{b^2+12 \pi  b+32 \pi ^2}  \geq 0.
\end{multline}
\end{widetext}

The energy density and EoS parameter for this case are depicted below.
\begin{figure}[H]
\centering
\includegraphics[width=7.5 cm]{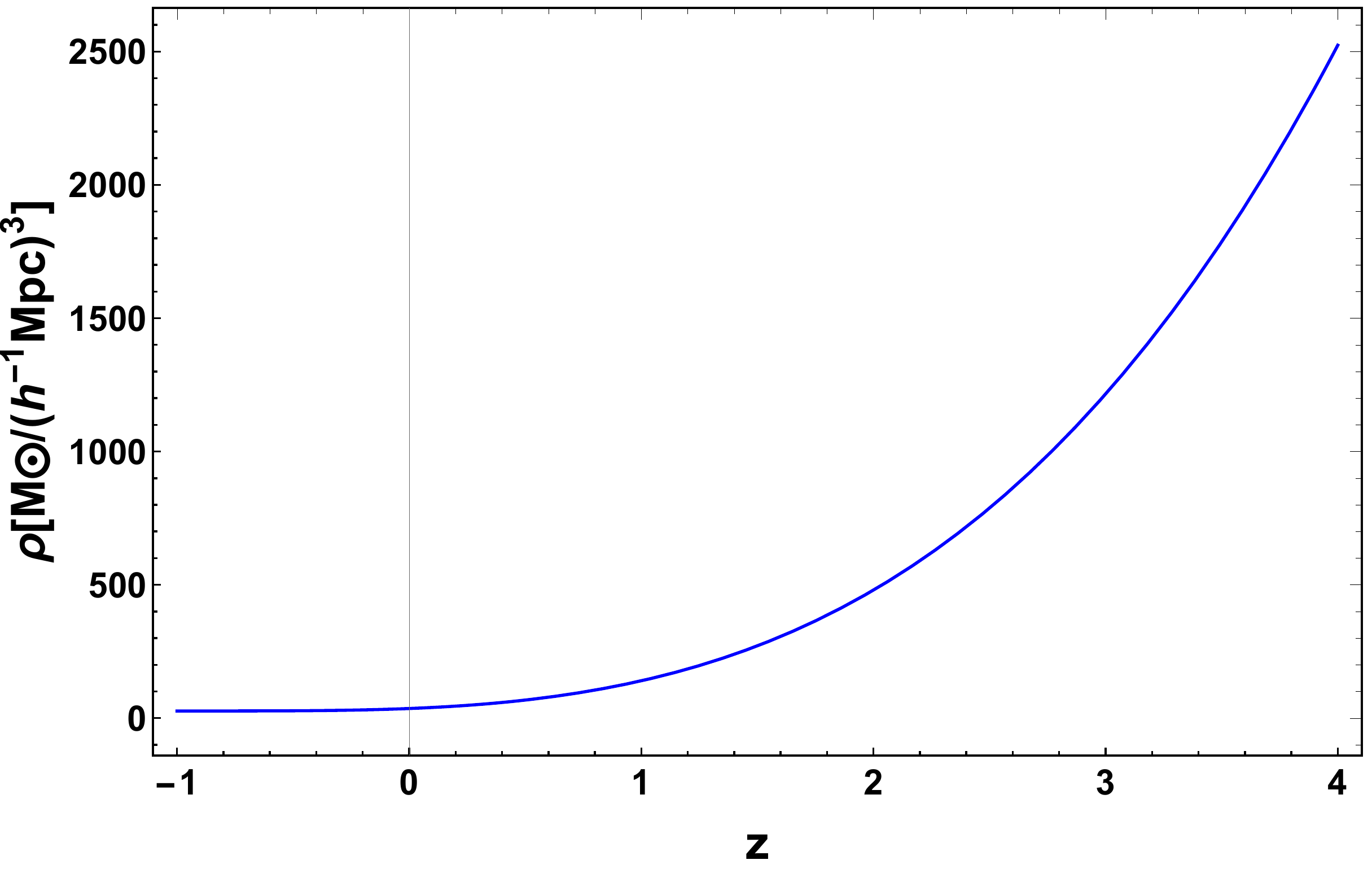}
\caption{Density parameter versus redshift.}\label{Fig-density2}
\end{figure} 

\begin{figure}[H]
\centering
\includegraphics[width=7.5 cm]{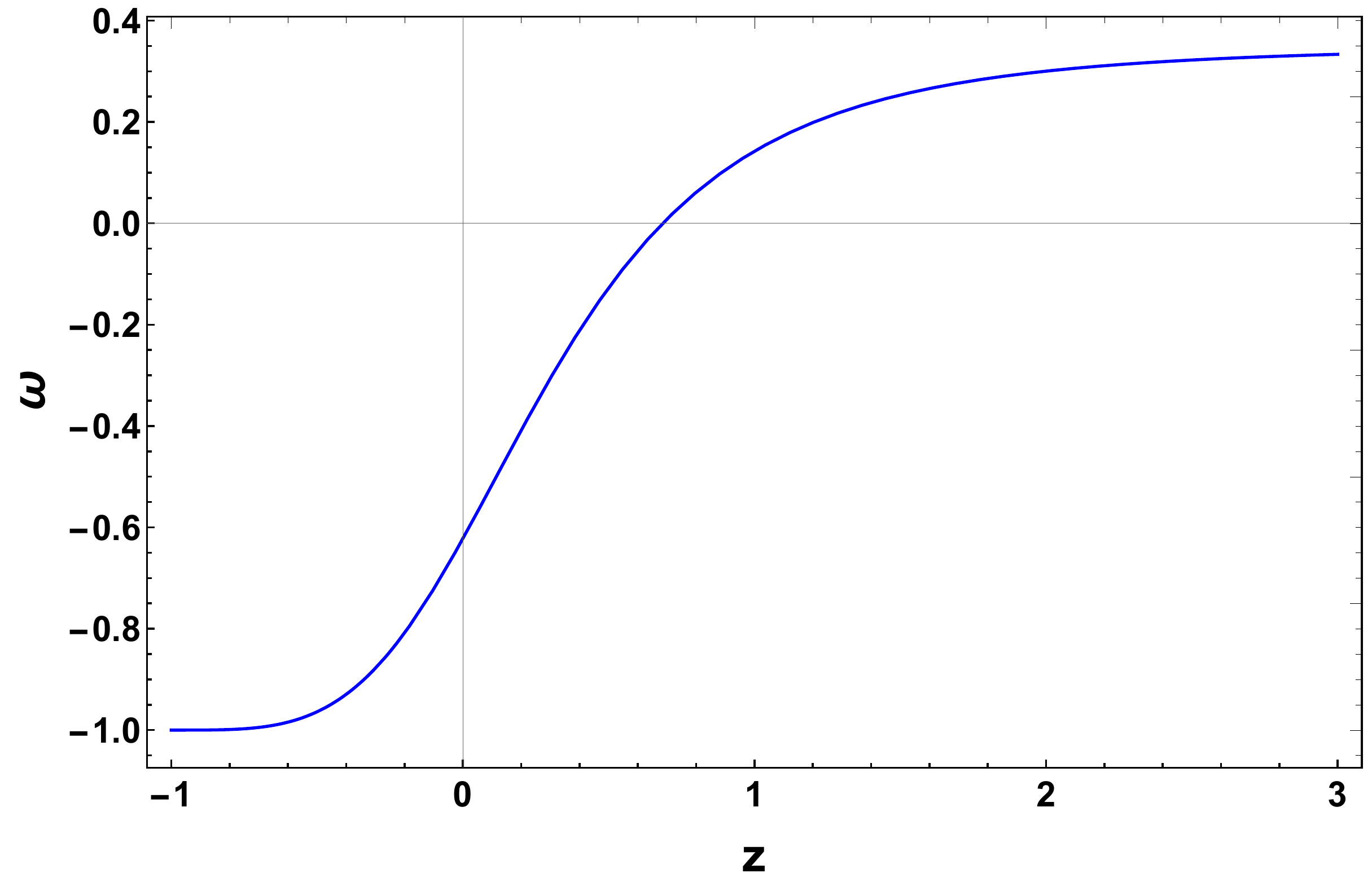}
\caption{EoS parameter versus redshift.}\label{Fig-Omega2}
\end{figure} 

Fig. \ref{Fig-density2} shows the behavior of energy density with respect to redshift with appropriate choice of parameters as $m=-0.1, b=59.1, n=0.1, q_{1}=3.57, q_{2}=3.82$ and $H_{0}=68.9$.The energy density can always be observed to be a positive redshift measure. At $z=0$, the energy density is strictly positive and increases with the increase in the value of $z$. By Fig. \ref{Fig-Omega2}, we can observe that the Universe has a transition from early deceleration to late-time accelerated phase at around $z=0.685$.

The graphical representation of energy conditions are given below.
\begin{figure}[H]
\centering
\includegraphics[width=7.5 cm]{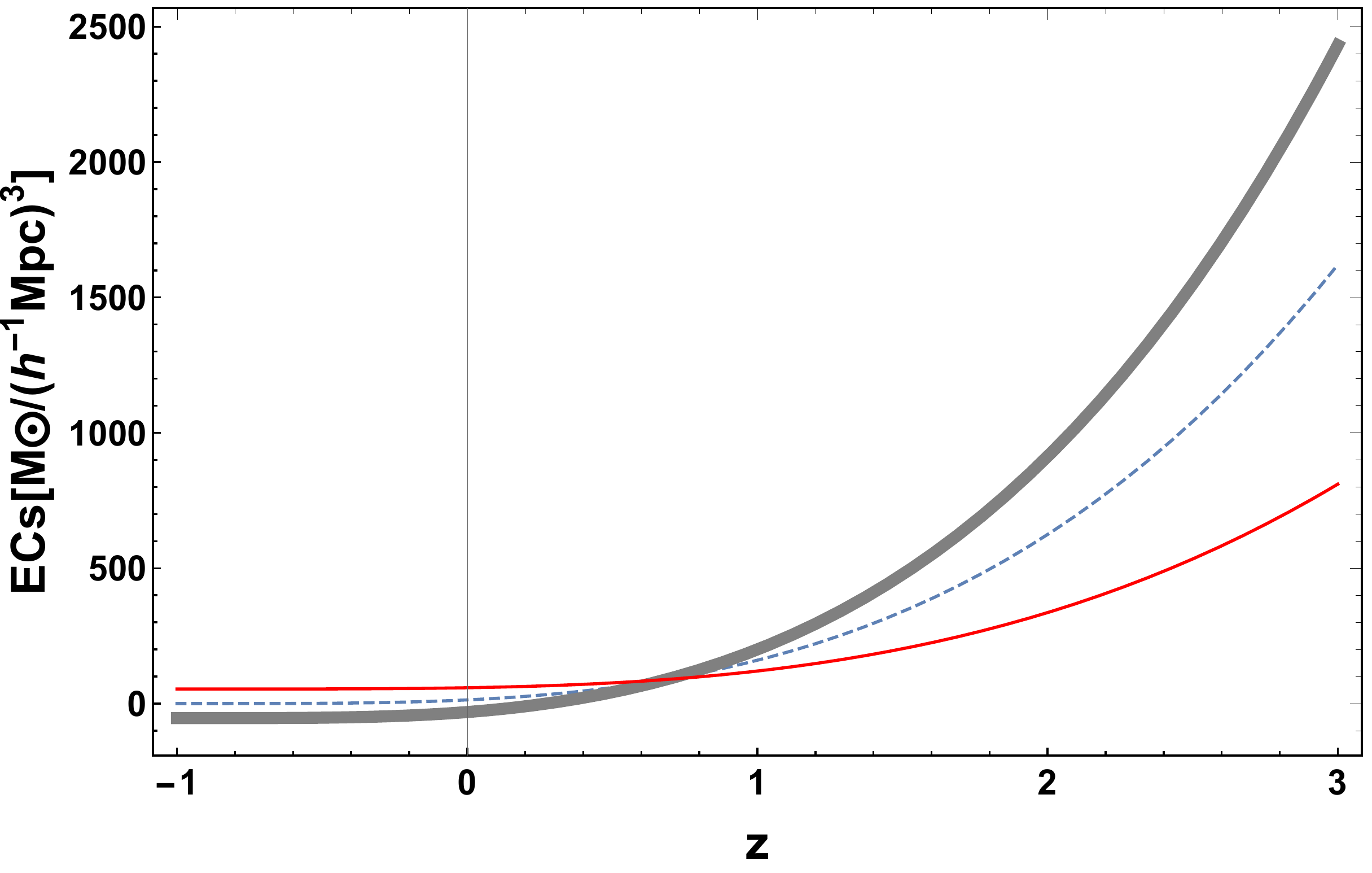}
\caption{Energy conditions versus redshift.}\label{Fig-EC2}
\end{figure} 

\begin{widetext}
\begin{figure}[H]
  \centering
  \subfloat[variation of b]{\label{SEC2b}\includegraphics[width=70mm]{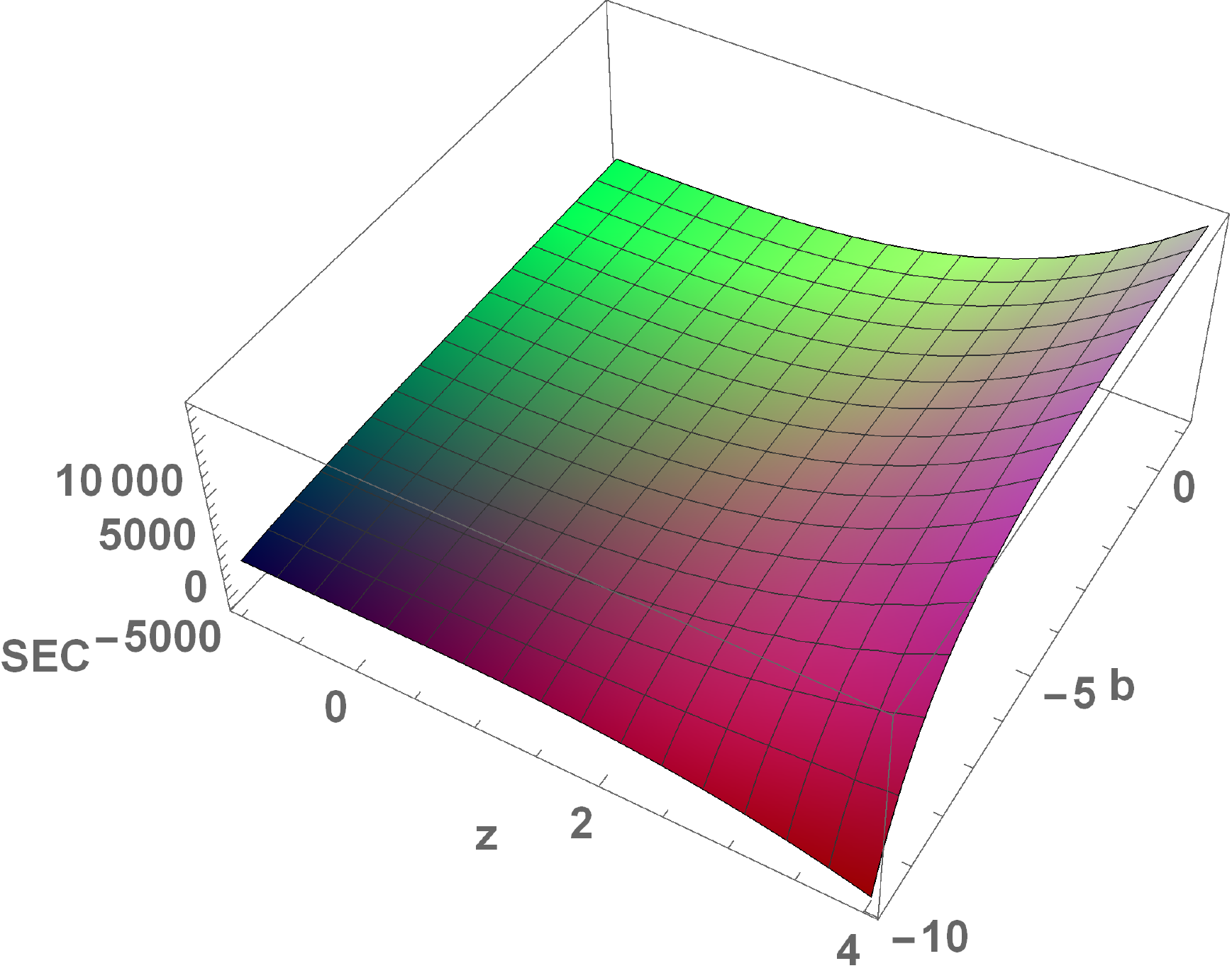}}
  \subfloat[variation of m]{\label{SEC2m}\includegraphics[width=70mm]{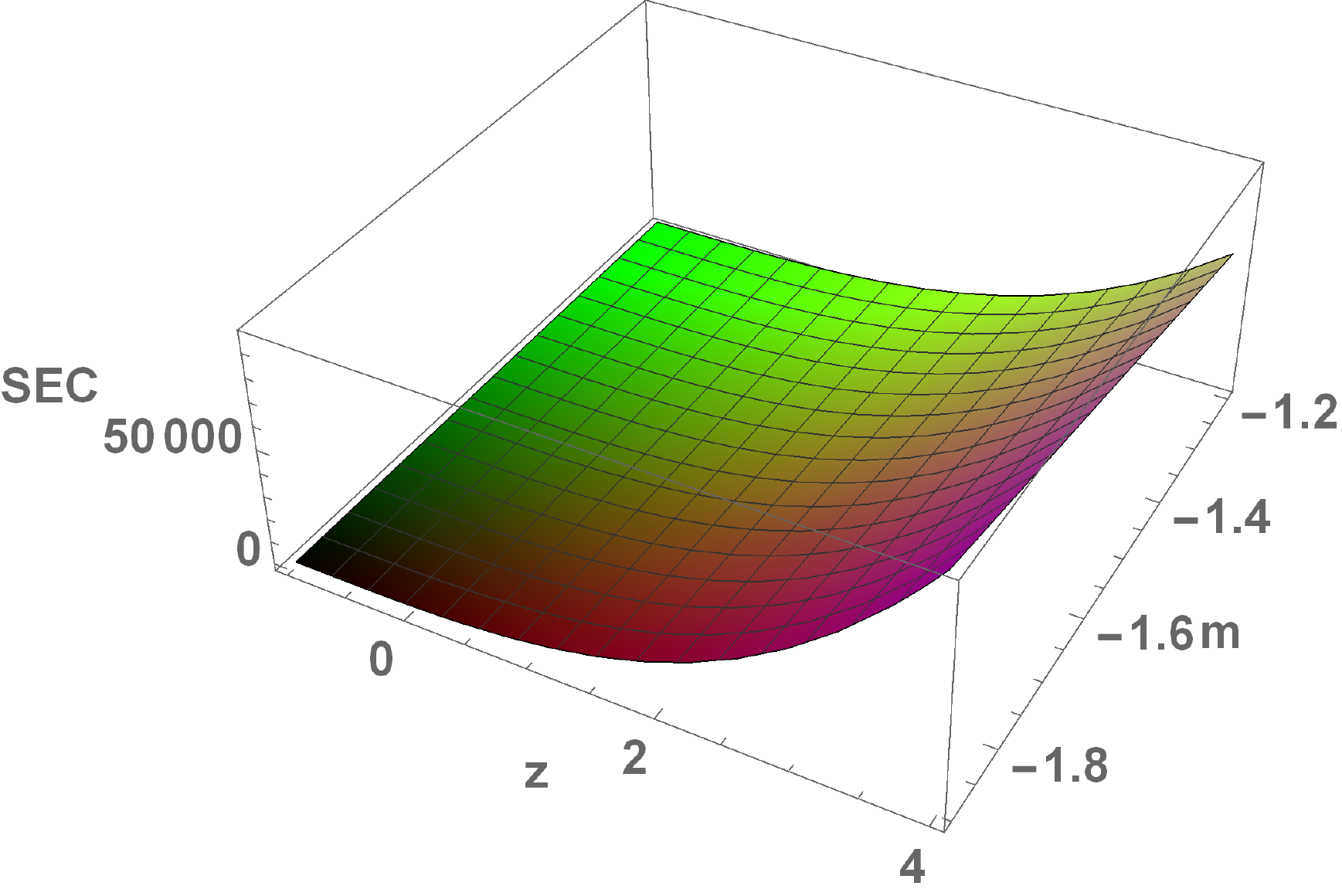}}
  \caption{ SEC with variation of b and m.}
\end{figure} 
 
\begin{figure}[H]\label{Fig-NEC2}
  \centering 
  \subfloat[variation of b]{\label{NEC2b}\includegraphics[width=70mm]{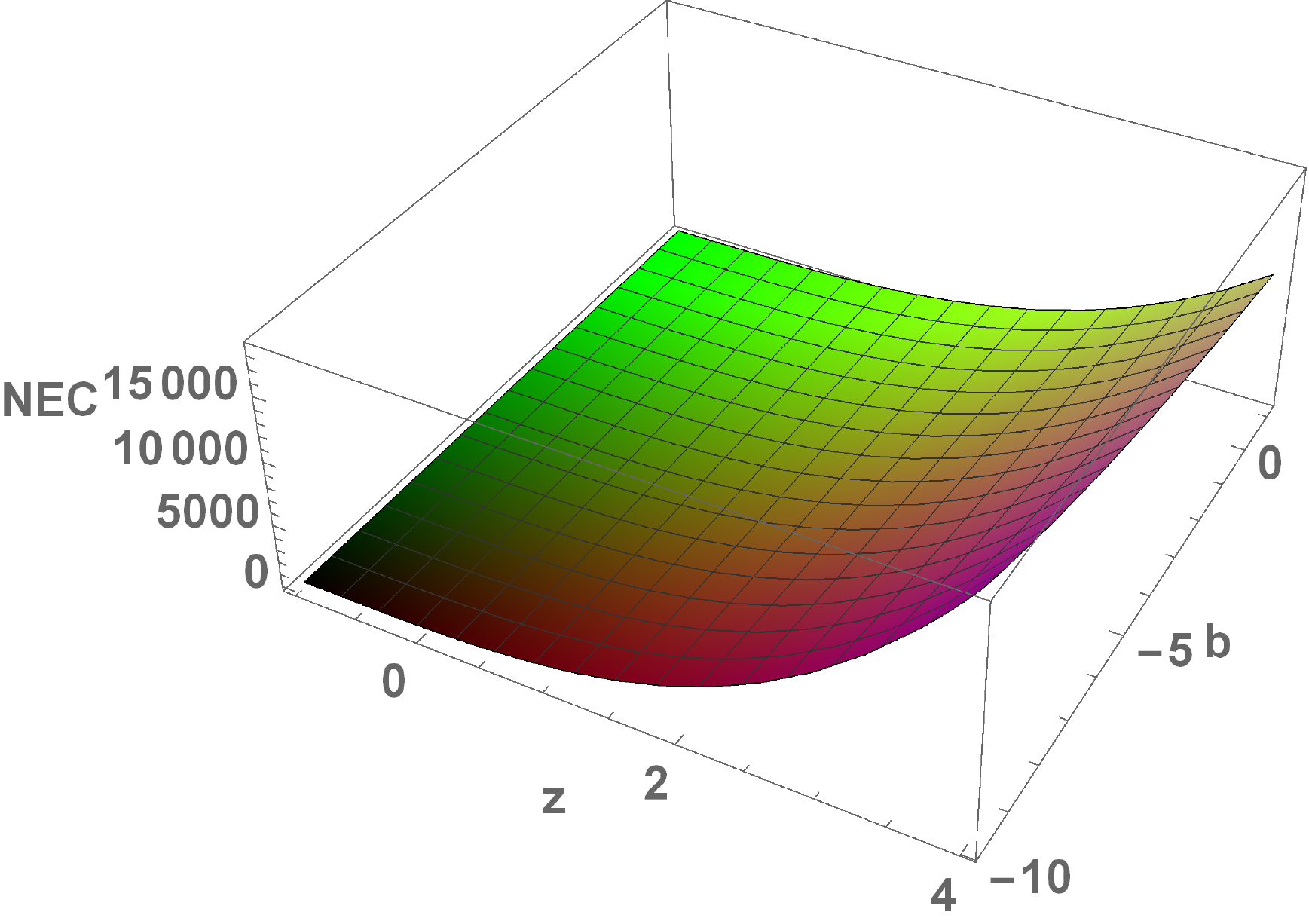}}
  \subfloat[variation of m]{\label{NEC2m}\includegraphics[width=70mm]{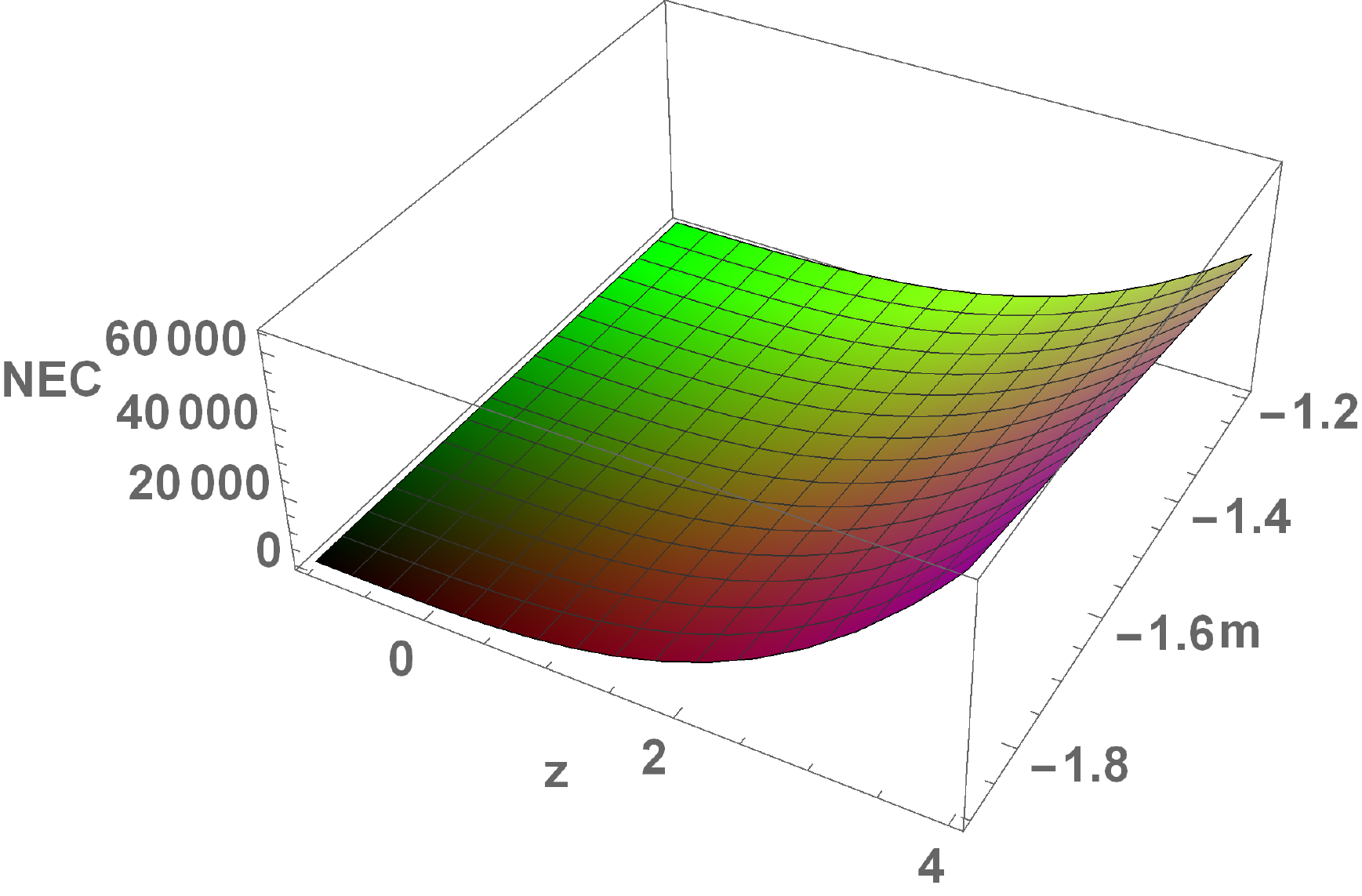}}
  \caption{ NEC with variation of b and m.}
 \end{figure} 
\end{widetext}

\begin{widetext}
\begin{figure}[H]\label{Fig-DEC2}
  \centering 
  \subfloat[variation of b]{\label{DEC2b}\includegraphics[width=80mm]{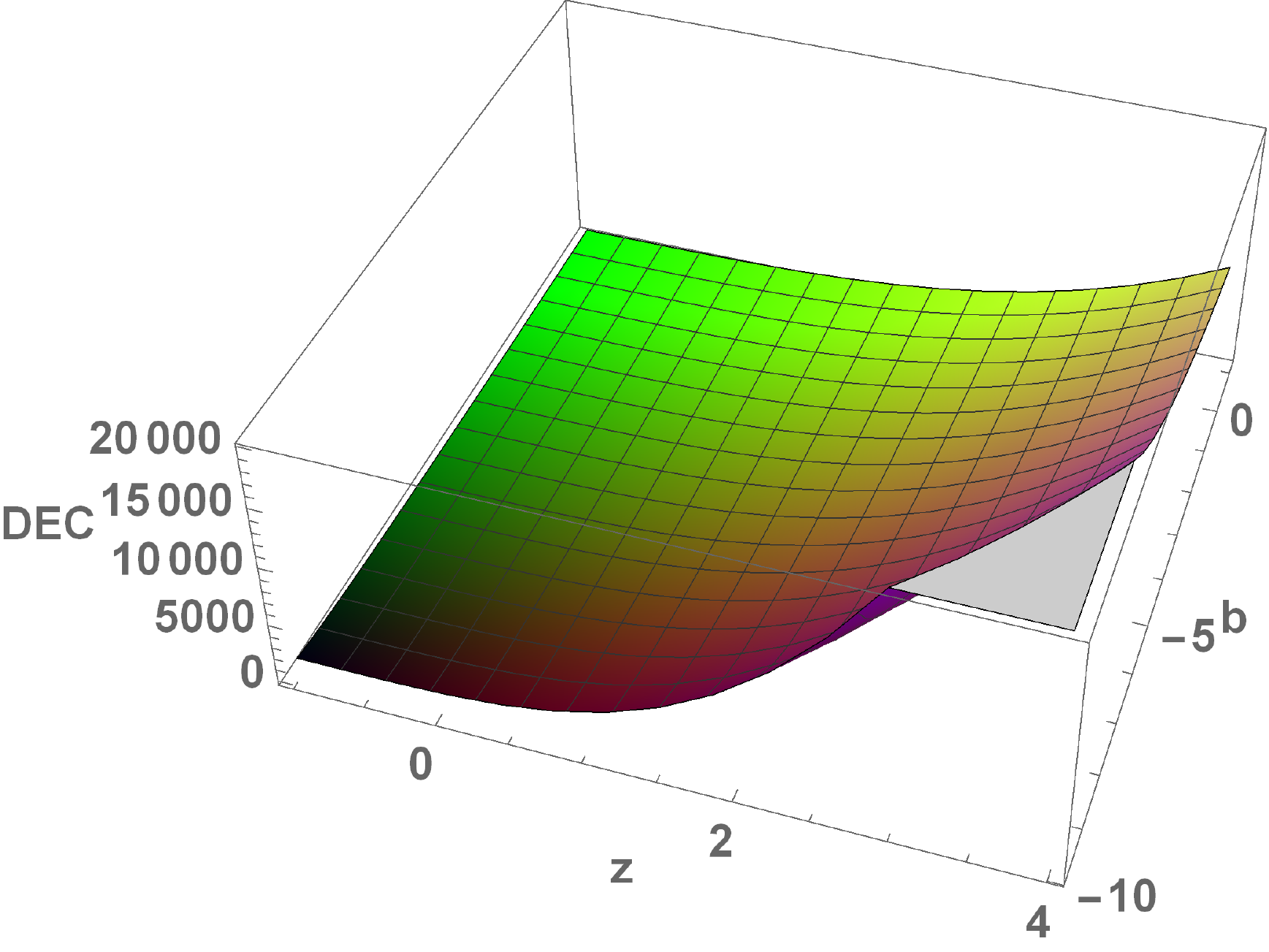}}
  \subfloat[variation of m]{\label{DEC2m}\includegraphics[width=80mm]{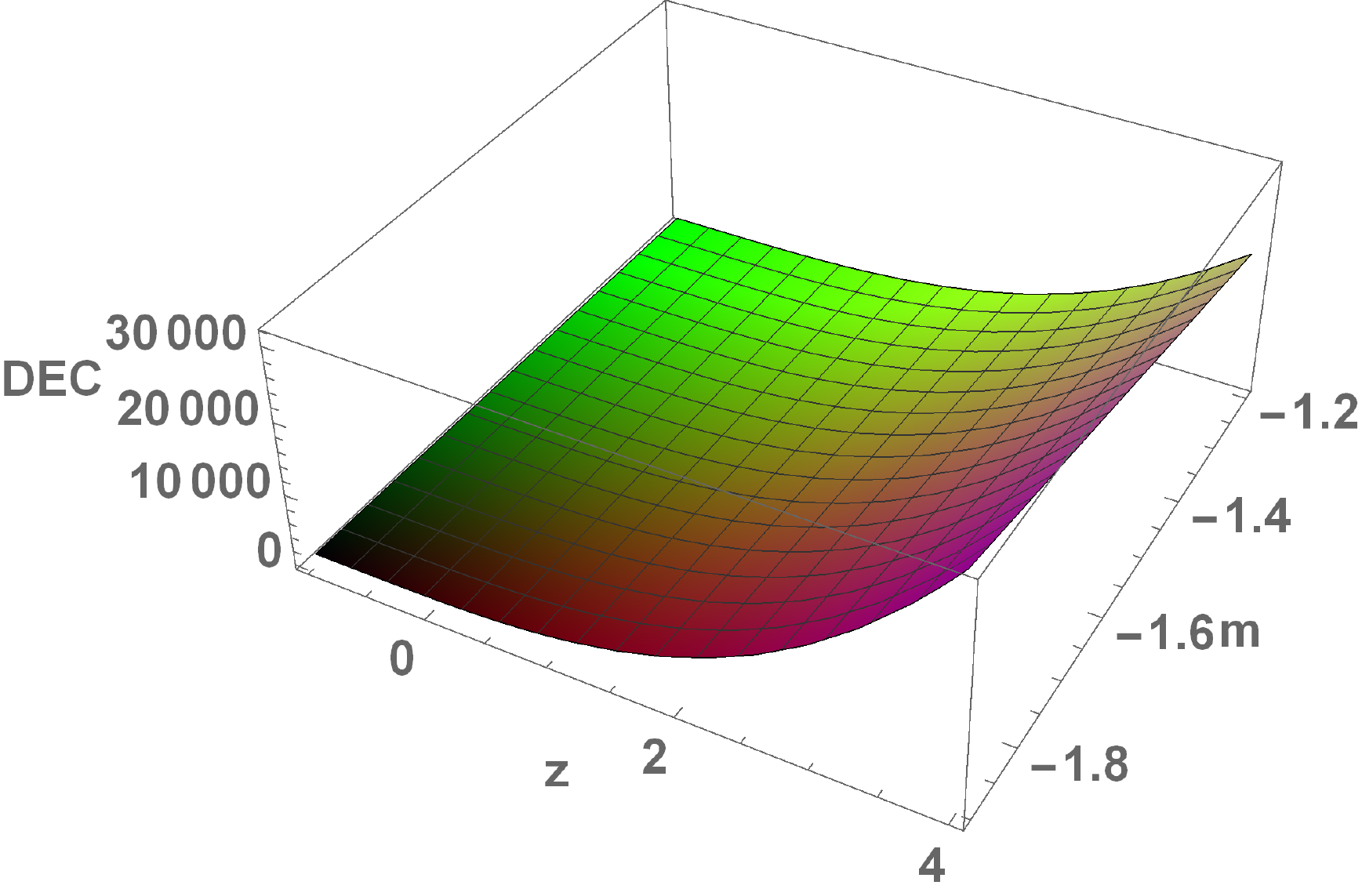}}
\caption{ DEC with variation of b and m.}
\end{figure} 
\end{widetext}

In Fig. \ref{SEC2b} and Fig. \ref{SEC2m} we plotted SEC by varying $b$ and $m$ respectively.  The variation of $m$ is from $-1.2$ to $-1.9$ whereas $b$ ranges from $-9.8$ to $1.1$. Therefore, there is a violation of the SEC at the present epoch with the variation of $b$. The slight change in the values of $ b $ towards the negative side results in the SEC's change. SEC violates more in the range mentioned above.  We can see that the NEC, and DEC do not violate as the behavior is always positive (see Fig. \ref{Fig-EC2},\ref{NEC2b},\ref{NEC2m},\ref{DEC2b},\ref{DEC2m}). Therefore, validation of NEC and energy density together results in the validation of WEC. The negative behavior of SEC illustrates the accelerated expansion of the Universe. 

\section{Conclusion}\label{sec6}

There has been a motivation by a recent observational study that the Universe is undergoing an accelerated expansion. This expansion is due to the high negative pressure known as ``dark energy" satisfying $\rho+3P<0$. Recently, a study on the geometrical extension to GR was known as modified theories of gravity.

In this paper, we have considered the energy conditions in recently proposed $f(Q,T)$ gravity. We used a proposed deceleration parameter with two unknowns $q_{1}$ and $q_{2}$ to solve the field equations. We also studied the behavior of the deceleration parameter with the values of contestants as $q_{1}=3.57$, $q_{2}=3.82$, and $H_{0}=68.9$ as obtained from observational constraints with respect to redshift. The $q$ deceleration parameter shows the transition from early deceleration to the Universe 's current acceleration at $z_{t}=0.67$with $q_{0}=-0.67$in accordance with the previous values given in the literature. After the deceleration parameter, the late-time acceleration is illustrated by the jerk, snap, and lerk parameters.

Further, we considered two models of $f(Q, T)$ gravity in section \ref{sec5} and observed the behavior of density, EoS parameter, and energy conditions. The density in both models shows positive behavior, whereas the EoS parameter shows a transition from early deceleration to late time acceleration with respect to the model parameters. According to the behavior of the EoS parameter, values of model parameters $m$ and $b$ are also used to examine various energy conditions. Like the deceleration parameter, the EoS parameter has a transition from deceleration to acceleration at $z=0.835$ and $z=0.685$ respectively, as presented in \cite{Capozziello}. In both the models of section \ref{A} \& \ref{B}, NEC, WEC and DEC are satisfied whereas SEC is violated (see Fig. \ref{Fig-EC1} \& Fig. \ref{Fig-EC2}). In a similar line, one can verify that minimally coupled \& curvature coupled scalar field theories also violate SEC \cite{Barcelo/2002}. In the present models, we can interpret the prediction of cosmic acceleration through energy conditions. 

Here, we have studied the linear cases of $f(Q, T)$ gravity and the non-minimal coupling i.e., the third model in \cite{Yixin} can be studied in the future.

\section*{Acknowledgments}  S. A. acknowledges CSIR, Govt. of India, New Delhi, for awarding Junior Research Fellowship. PKS acknowledges CSIR, New Delhi, India for financial support to carry out the Research project [No.03(1454)/19/EMR-II Dt.02/08/2019]. 


\end{document}